# Skyrmion Based Magnonic Crystals

Zhendong Chen and Fusheng Ma[*]

*Jiangsu Key Laboratory of Opto-Electronic Technology, Center for Quantum Transport and Thermal Energy Science, School of Physics and Technology, Nanjing Normal University, Nanjing 210046, China*

*Correspondence to Fusheng Ma: phymafs@njnu.edu.cn

## Abstract

Magnonics is now an attractive field which focuses on the dynamic characteristics of magnons, a kind of quasiparticles in magnetic media, and attempt to apply magnons for functional devices. In order to construct magnon based devices, it is necessary to fabricate materials with specific and tunable magnon bands and band gaps. Skyrmion based magnonic crystals is one of the most suitable materials which possess periodical skyrmion structure and show applicative magnon bands and band gaps. In this review, we provide an overview over recent theoretical and experimental research on skyrmion based magnonic crystals. We will firstly provide an introduction of magnonic crystals and magnetic skyrmion. And then we will show the theoretical and experimental progress on skyrmion based magnonic crystals and their magnon band characteristics. At the end, we will give an outlook and perspectives of new fascinating fields on topological nontrivial magnon modes, as well as hybrid and quantum magnonic phenomena of skyrmion based magnonic crystals.

# 1. Introduction

The concept of spin waves (SWs) was firstly introduced by F. Bloch 80 years ago. From a view of classical theory, SWs represent phase-coherent precession of the microscopic vectors of magnetization in the magnetic medium, which can be considered as a magnetic analogue of a sound or light wave in corresponding media[1–4]. Magnons are the quanta of SWs, which should behave as weakly interacting quasiparticles obeying the Bose-Einstein statistics[5,6]. Many previous theoretical and experimental works have demonstrated that SWs or magnons exhibit most of the properties inherent in waves or quasiparticles of other origins. For example, the excitation and propagation[7–15], reflection and refraction[16–24], interference and diffraction[25–29], focusing and self-focusing[30–39], tunneling[40,41] of SWs, Doppler effect[42–44], formation of SW envelope solitons[45–48], and even Bose-Einstein condensation of magnons[49–51] were observed.

Such encouraging observations has stimulated a new attractive field of magnonics[1,52–61]. As an emerging subfield of spintronics[62,63], the purpose of magnonics is revealing the physical characteristics of SWs (or magnons), and using SWs to carry or process information on the microscale and nanoscale. Fig. 1 shows several active study fields of magnonics in recent years. From the physics point of view, many novel characteristics of magnons are reported beside the fundamental characteristics mentioned above, and lead to new subfields such as topological magnonics, hybrid magnonics, and quantum magnonics, *et al.*[60,64,65,66,67]. From the application point of view, there are many efforts been reported to achieve magnon-based data processing. Comparing with electron-based data processing, magnon-based data processing can show new advantages such as high efficiency, multifunctionality, tunability, broad frequency range, and low energy dissipation[59–61,68–71]. Several microscale and nanoscale devices have been designed such as logic gates, magnon transistors, interferometers and high frequency devices, *et al.*[59,72–84].

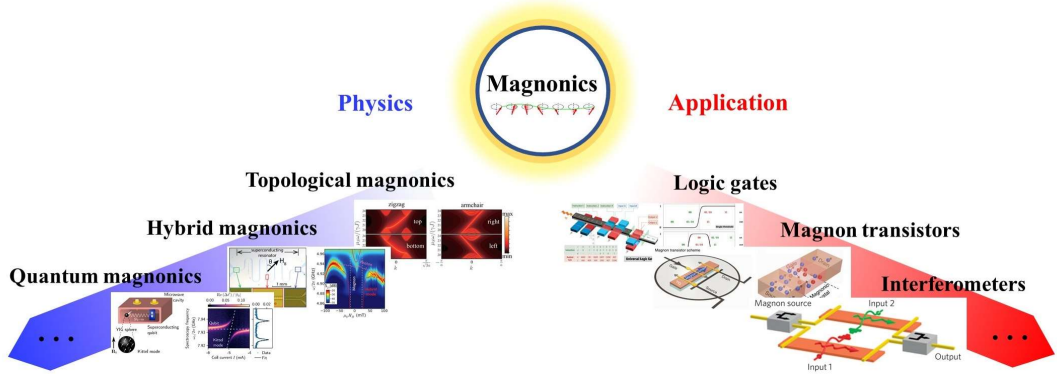

Fig. 1. Illustration of the active fields in magnonics. Such fields can be classified briefly into physics studies which concern the novel characteristics of magnons in condensed matter systems, and application studies which aim to apply magnons for information processing. [Reproduced with permission from X. S. Wang *et al.* J. Appl. Phys. 129, 151101 (2021), with the permission of AIP Publishing. Y. Li, *et al.* Phys. Rev. Lett. 123, 107701 (2019). Copyright 2019, American Physical Society. D. Lachance-Quirion, *et al.* "Hybrid quantum systems based on magnonics", Appl. Phys. Express Vol. 12, 70101, 2019, DOI: 10.7567/1882-0786/ab248d; licensed under a Creative Commons Attribution 4.0 license. W. Yu, *et al.* Phys. Rev. Appl. 13, 024055 (2020). Copyright 2020, American Physical Society. A.V. Chumak *et al.* Nat. Commun. Vol. 5, 4700, 2014; licensed under a Creative Commons Attribution-NonCommercial-ShareAlike 4.0 International License. T. Schneider, *et al.* Appl. Phys. Lett. 92, 022505 (2008), with the permission of AIP Publishing.]

Further studies of magnonics and magnon-based devices require a platform in which different magnon modes can be excited. In other words, studies of magnonics and magnon-based devices call for a new kind of magnetic media which possess specific magnon band structures with "valence" and "conductive" bands as well as band gaps. Thus, the concept of magnonic crystals (MCs) was put forward[61,69,70,85–87], in which spin waves (or magnons) can be modulated into magnon bands via artificial periodical structures of magnetic properties. In such a new kind of magnetic media, skyrmion based MCs (SMCs) have attracted much attention due to their dynamically tunability, topological nontrivial characteristics, as well as hybrid and quantum phenomena[88–106]. In this review, we will firstly introduce the concepts of MCs and magnetic skyrmion. Then we will show recent theoretical and numerical studies on magnon band characteristics of one-dimensional SMCs, two-dimensional SMCs, antiferromagnetic SMCs, antiskyrmion MCs, as well as the experimental efforts to investigate the dynamic properties of skyrmion crystals and achieve artificial skyrmion lattices. At the end, we will give an outlook and perspectives of new fascinating fields on using SMCs to investigate topological magnonics, hybrid magnonics, and quantum magnonics.

## 2. What are magnonic crystals?

MCs represent a kind of artificial magnetic media with specific magnon properties characterized by periodical magnetic structures. The spectra of SW excitations in such magnetic media are significantly different from those in uniform media: the SW spectra of MCs exhibit features such as magnon bands where spin waves can propagate easily, as well as band gaps where spin waves are not allowed to propagate[1,61,69,70,85–87]. Early studies on SWs in media with periodical magnetic structures began in the 1970s, which mainly focused on the development of microwave filters and resonators[107,108]. In 1990s, researchers started to reveal the unique characteristics of the photon/phonon/magnon spectra with band gaps in the artificial periodic structures of corresponding media, which are called photonic/phononic/magnonic crystals, respectively[52,109,110]. Since then, MCs are of great interest in both pure wave physics and application oriented magnonics[58–60,69,70]. Fig. 2 shows several typical periodical structures of one-dimensional (1D), two-dimensional (2D), and 3-dimensional (3D) MCs and their magnon bands, respectively[87,88,111–120]. These periodical modulations of magnetic structures can be achieved by both the external geometrical modifications and the intrinsic properties of the magnetic media[88,111–118]. The geometrical modifications can be realized by chemical etching, laser processing or ion implantation, *et al.*[85–87,111–114]. While the intrinsic properties can be introduced by the local internal magnetic fields, local magnetic states or local spin textures[88,115–118]. Nowadays, many special SW-related physical phenomena have also been observed in MCs, such as topological and nonreciprocal phenomena[92–98,121–125], linear and nonlinear spin-wave dynamics in coupled MCs[126–128], and the formation and propagation of solitons[129–132].

In periodic media, waves in general undergo two opposite microscopic mechanisms for band formation[86]. On the one hand, waves can undergo coherent scattering and Bragg reflection in a periodically modulated material, giving rise to a modified band structure with partial or complete band gaps. On the other hand, starting from confined or standing modes in individual resonators, coherent coupling between them allows the creation of dispersive bands. The discrete eigenfrequencies $f$ transform into dispersion relations $f(k)$ that support waves propagating if arranged on an appropriate lattice ($k = 2\pi/\lambda$ is the wavevector). The eigensolutions of the wave equation fulfil Bloch's theorem in either case and, regardless of the type of excitation, form the band structure in reciprocal space. The artificially tailored band structure can be modulated by the geometrical parameters. As one kind of waves in magnetic media, SWs exhibit some unique $f(k)$ dispersion relations which also depend on some external effect such as the magnitude and direction of the magnetic field. Thus, the magnonic properties of the MCs can be flexibly controlled by more diversified methods compared with other artificial crystals.

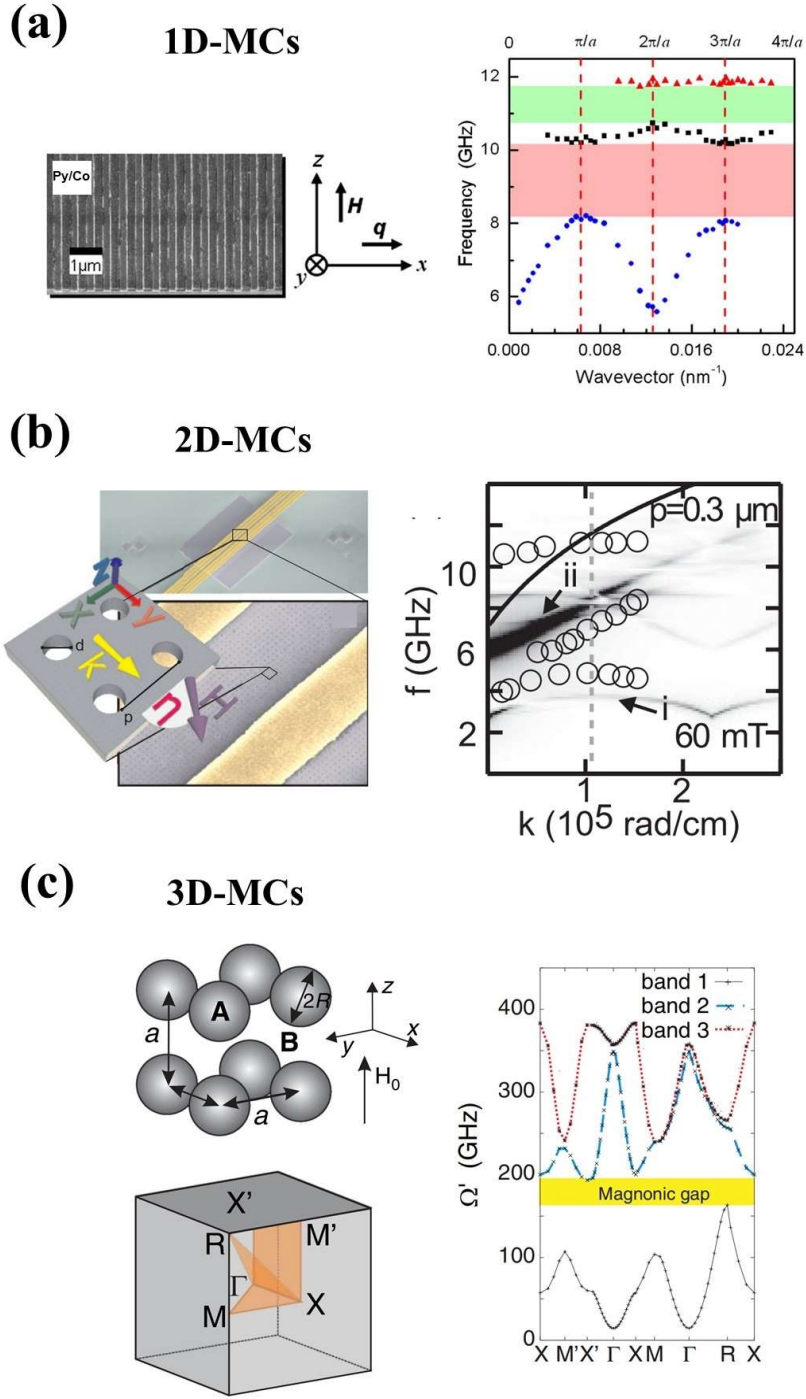

Fig. 2. Illustration of (a) 1D-, (b) 2D- or (c) 3D-MCs and their magnon band structures[87,119,120]. [(a) is reproduced from Z. K. Wang, *et al.* Appl. Phys. Lett. 94, 83112 (2009), with the permission of AIP Publishing; (b) is reproduced with the permission from S. Neusser, *et al.* Phys. Rev. B 84, 094454 (2011). Copyright 2012, American Physical Society; (c) is reproduced with the permission from J. Romero Vivas, *et al.* Phys. Rev. B 86, 144417 (2012). Copyright 2012, American Physical Society.]

# 3. What are magnetic skyrmions?

The magnetic skyrmion is a local whirl of the spin configuration in a magnetic material. It can be considered as a new type of magnetic quasi-particle in a magnetic medium[133–135]. As shown in the top panel of Fig. 3(a)[134], the spins inside a skyrmion rotate progressively with a fixed chirality from the up direction at one edge to the down direction at the center, and then to the up direction again at the other edge. There are two typical types of magnetic skyrmions: Néel-type and Bloch-type skyrmions, which correspond to different symmetries of the interaction between spins (this can be due to the underlying crystal lattice or to the presence of an interface, for instance), resulting in different directions of the rotation. In most systems, the spin configuration of skyrmions is determined by chiral interactions of the Dzyaloshinskii–Moriya type which is called Dzyaloshinskii–Moriya interactions (DMIs). The sketch of DMIs is exhibited in the bottom panel in Fig. 3(a). Such interactions are induced by the spin-orbit coupling in the systems with lack of inversing symmetry, and the Hamiltonian is written as[134]:

$$H_{DMI} = (S_1 \times S_2) \cdot d_{12} \tag{1}$$

where $S_1$ and $S_2$ are neighbouring spins and $d_{12}$ is the corresponding Dzyaloshinskii–Moriya vector. For the interfacial DMI in thin films, $d_{12}$ can be written as $d_{12} = d_{12} \cdot (z \times u_{12})$, where $z$ and $u_{12}$ are unit vectors, respectively perpendicular to the interface in the direction of the magnetic layer and pointing from site 1 to site 2. It can be easily pointed out that DMIs tend to make the two spins be arranged perpendicularly. The DMI is a chiral interaction that decreases or increases the energy of the spins depending on whether the rotation from $S_1$ to $S_2$ around $d_{12}$ is in the clockwise ($d_{12} < 0$) or in the anticlockwise ($d_{12} > 0$) sense. If $S_1$ and $S_2$ are initially parallel, the effect of a strong DMI (compared with the symmetric exchange interaction) is to introduce a relative tilt around $d_{12}$.

Skyrmions can, indeed, be defined by the topological number $S$ (or skyrmion number), which is a character of the winding of the normalized local magnetization, $m$.

In the two-dimensional limit, the topological number is[134]:

$$S = \frac{1}{4\pi} \int \boldsymbol{m} \cdot (\partial_x \boldsymbol{m} \times \partial_y \boldsymbol{m}) \mathrm{d}x \mathrm{d}y \qquad (2)$$

The normalized magnetization can be mapped on a unit sphere and, in the case of skyrmions, it covers the entirety of the sphere ($4\pi$) and is thus quantized. This non-trivial topology governs some of the most important properties of skyrmions such as the topological protection of the spin configuration and the quasi-particle properties[134,136].

The concept of skyrmions was firstly proposed in nuclear physics by T. H. R. Skyrme's original work in 1962[137]. Then in 1980s, the concept was extended into condensed matter physics[138,139]. Magnetic skyrmions were firstly predicted existing in magnetic media with non-centrosymmetric lattice by U. K. Rößler *et al.* in 2006[140], and then they were observed experimentally by S. Mühlbauer *et al.* via neutron scattering technique in 2009[141]. Since then, researchers have already found skyrmions in variety of bulk magnetic materials and magnetic thin films[133–135]. Because of their topological protection, small spatial size and low driving current density[136,142–147], magnetic skyrmions show remarkable potential applications in new high-density and low-energy-dissipation spintronic devices such as racetrack memory, logic devices and radio-frequency devices[114,157–161]. In order to achieve the goal of applying skyrmions to these devices, researchers have invested much attention on the thermal stability, nucleation, annihilation, detection and current-driven movement in recent years[133–135]. Since 2015, researchers combined the concept of skyrmions and MCs, and predicted a new kind of MCs which called SMCs[88–100,104,106]. Fig. 3(b) exhibit the illustrations and the magnon band structures of 1D- and 2D-SMCs. It should be pointed out that the periodical magnetic structures of SMCs are constituted by skyrmions. And the periodical skyrmion arrangement can also induce the generation of magnon bands as well as band gaps. Recent reports revealed that SMCs can provide topological nontrivial magnon modes and chiral edge states, as well as hybrid magnonic phenomena and quantum phenomena. The details of the recent achievements and perspective of SMCs will be presented in the following.

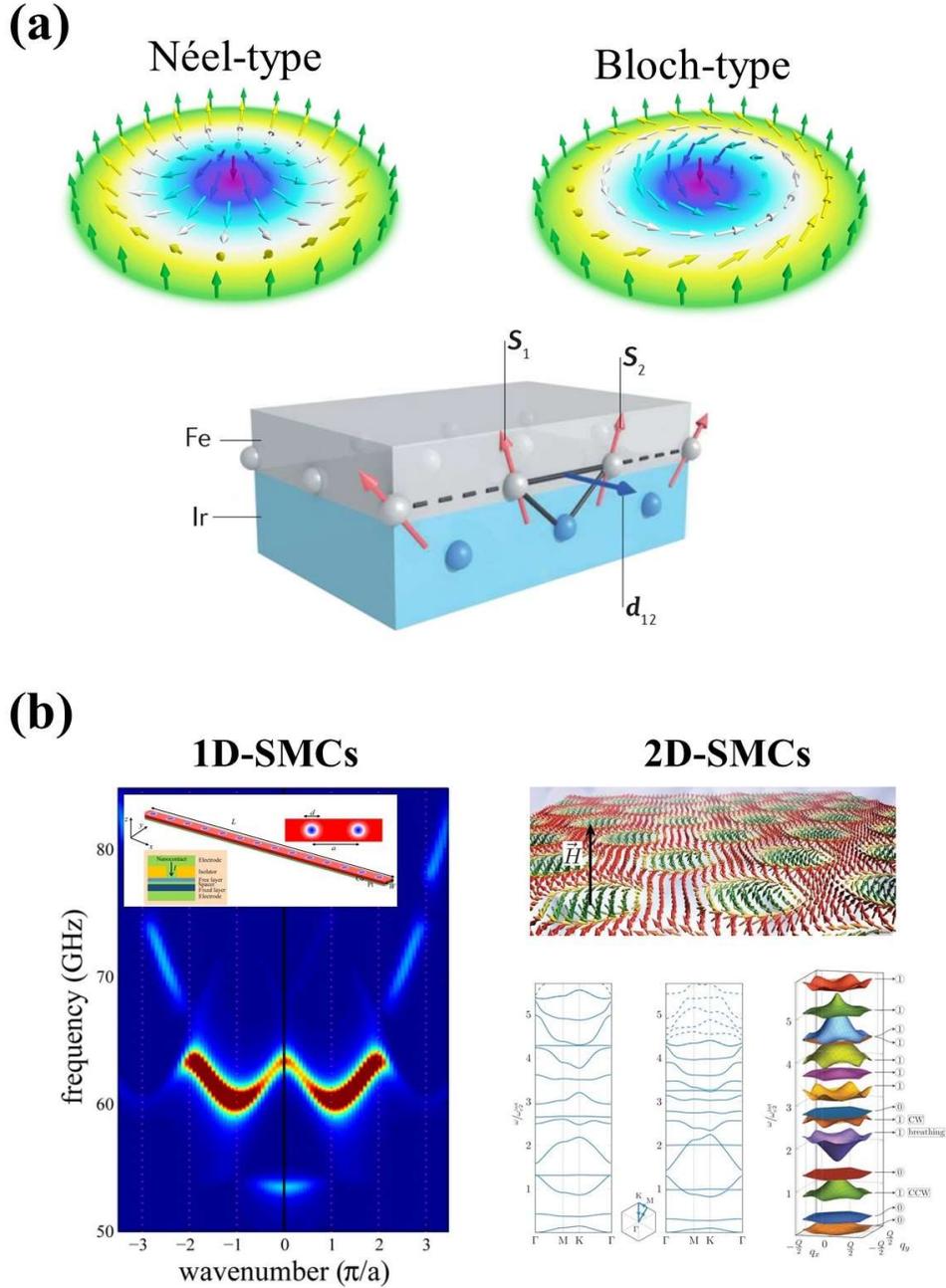

Fig. 3. (a) Spin texture of skyrmions. Néel- and Bloch-type skyrmions and the sketch of DMI in Fe/Ir bilayer[134,148]. (b) The sketch and the magnon bands of 1D SMC (left panel) and 2D SMC (right panel), respectively[88,100]. [(a) is reproduced with the permission from A. Fert, *et al.* Nat. Nanotechnol. 8, 152 (2013). Copyright 2013, Springer Nature. (b) is reproduced with the permission from F. S. Ma, *et al.* Nano Lett. 15, 4029 (2015). Copyright 2015, American Chemical Society; M. Garst, *et al.* "Collective spin excitations of helices and magnetic skyrmions: review and perspectives of magnonics in non-centrosymmetric magnets", J. Phys. D: Appl. Phys. Vol. 50, 293002, 2017, DOI: 10.1088/1361-6463/aa7573, licensed under a Creative Commons Attribution CC-BY 3.0 license.]

# 4. What are skyrmion based magnonic crystals?

SMCs are a new style of MCs in which the artificial periodic structures are constituted by skyrmions[88–100,104,106]. Because the presence, size and site of the skyrmions can be modulated by external effects, for example, external magnetic fields, electric currents, electric fields or spin waves[148,154–158], the magnon band structure of SMCs, especially the existence of magnon band gaps, can be modulated by such external effects while operating. This advantage makes SMCs to be classified into a new concept of "dynamic magnonic crystals (DMCs)", which means these MCs possess tunable magnon band structures and switchable band gaps after they have been already prepared[88,159,160]. More interestingly, topological magnons, chiral magnonic edge state, as well as hybrid and quantum magnonic phenomena were predicted in SMCs in recent works[92–98,103,106]. According to the spatial dimension of the periodical skyrmion arrangement, SMCs can be classified into two types: 1D-SMCs[88–91] and 2D-SMCs[92–99]. Typical theoretical and numerical studies on these two classes of SMCs will be introduced in this section. In recent years, there are other special kinds of 2D-SMCs been predicted theoretically which are constituted by other general skyrmion-like spin textures, such as antiferromagnetic skyrmions and antiskyrmions. These two types of SMCs will be individually introduced in this section[95,98]. Beside the theoretical and numerical studies on SMCs, we will also review experimental achievements on the dynamic modes of skyrmion crystals, and building artificial skyrmion crystals, both of which can promote the application of SMCs for future magnonic devices.[100–103,161–166].

**4-1. One-dimensional skyrmion based magnonic crystals**

Since 2012, there were several works discussing the interaction between spin waves and single skyrmion or skyrmion lattices[161,167–170]. However, the concept of SMCs was firstly come into people's view along with the 1D-SMC was put forward[88]. In 2015, F. S. Ma *et al.* firstly reported a numerical study on 1D-SMCs which based on a perpendicularly magnetized Co/Pt waveguide with the presence of interfacial DMI. The magnetic periodicity of the waveguide is realized by the presence of a spatially periodic array of skyrmions created by nanocontacts carrying a spin current. The presence or absence of the skyrmions can be dynamically controlled by external magnetic fields or the spin current in the nanocontacts. Hence, the artificial skyrmion lattice of the SMC can be switched from "off" to "on" and make the spin wave modes lying within the bandgap become the forbidden modes (and vice versa). Correspondingly, the spin wave transmission is switched from full transmission to full rejection, which is exhibited in Fig. 4(a). In 2016, M. Mruczkiewicz *et al*. theoretically investigated the magnetization dynamics in an 1D-SMC[89]. They reported that the determined dispersion relation of coupled skyrmions shows a characteristic feature of the band structure in MCs, and they identified the excited modes in the magnon bands as breathing and clockwise gyrotropic dynamic skyrmions, which correspond to the high- and low-frequency modes, as shown in Fig. 4(b). In 2017, J. Kim *et al*. investigated the coupled gyrotropic modes in 1D-SMCs induced by the interaction between neighboring skyrmions, as shown in Fig. 4(c). They also indicated that these modes can be modulated by the distance between the neighboring skyrmions and the perpendicular external magnetic fields[90]. Then in 2018, they subsequently reported the coupled breathing modes in 1D-SMCs, and identified the in-phase high-energy mode at zero wavenumber and the anti-phase low-energy mode at the Brillouin zone boundary, as exhibited in Fig. 4(d)[91].

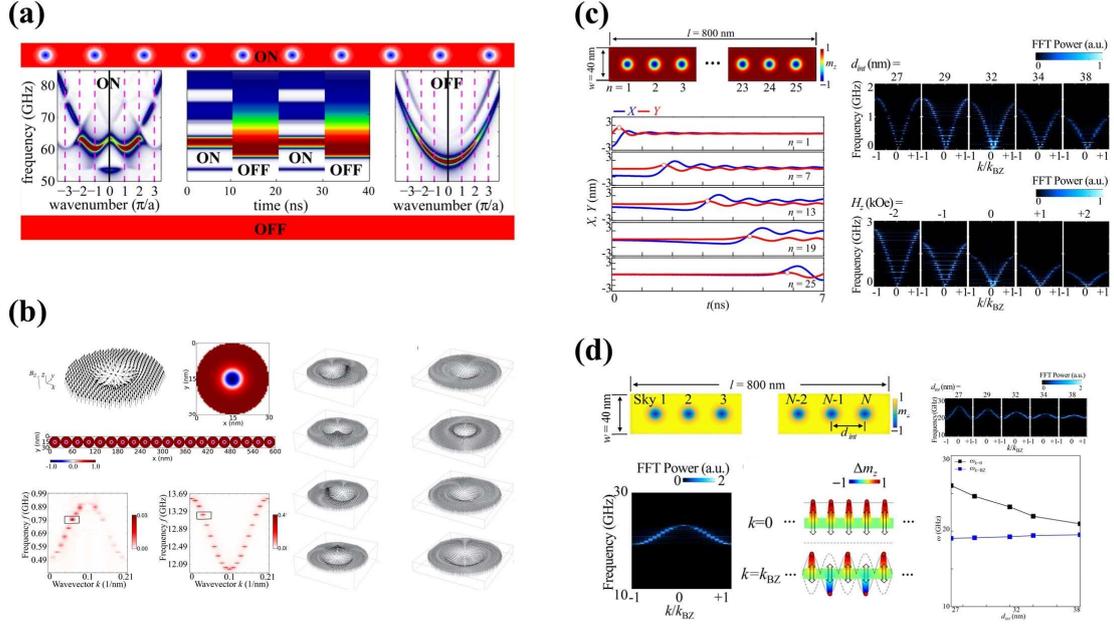

Fig. 4. (a) Switching on and off of the 1D-SMC's skyrmion lattice and the magnon band structures[88]. (b) The sketch of the 1D-SMC, the magnon band structures of the clockwise gyrotropic and breathing skyrmion excitation modes, and the real spatial profiles of the two excitation modes[89]. (c)-(d) The sketch of the coupled 1D-SMC, the real spatial profiles and the magnon band structures of the coupled gyration modes (c) and the coupled breathing modes (d), respectively[90,91]. [(a) is reproduced with the permission from F. S. Ma, *et al.* Nano Lett. 15, 4029 (2015). Copyright 2015, American Chemical Society. (b) is reproduced with the permission from M. Mruczkiewicz, *et al.* Phys. Rev. B 93, 174429 (2016). Copyright 2016, American Physical Society. (c) is reproduced with the permission from J. Kim, *et al.* Sci. Rep. Vol. 7, 45185, 2017; licensed under a Creative Commons Attribution CC-BY 4.0 license. (d) is reproduced from J. Kim, *et al.* J. Appl. Phys. 123, 053903 (2018), with the permission of AIP Publishing.]

**4-2. Two-dimensional skyrmion based magnonic crystals**

The concept of 2D-SMCs is a spatial extension of 1D-SMCs, which means SMCs with two dimensional periodic skyrmion arrays. 2D skyrmion lattice has already been observed in $Fe_{0.5}Co_{0.5}Si$ thin film in 2010 [171]. In 2012, M. Mochizuki theoretically raised three collective dynamic modes of skyrmion lattice: clockwise, counterclockwise and breathing modes[170]. Since then, these collective dynamic modes were experimentally investigated in several natural skyrmion-lattice systems such as $Cu_2OSeO_3$, MnSi, and $Fe_{0.8}Co_{0.2}Si$[100–103,161]. However, the potential of 2D skyrmion lattices for new kind of MCs is recognized after the 1D-SMCs were reported. In 2015,

A. Roldán-Molina *et al.* firstly showed a numerical result on the magnon bands of the 2D-SMCs, as shown in Fig. 5(a)[104]. Then in 2016, they calculated the topological properties for spin waves in 2D triangular skyrmion crystal based on Fe/Ir multilayers[92]. It was revealed that these bands have a finite Berry curvature and non-zero Chern number. Thus, a 2D skyrmion lattice would realize the spin-wave analogue of the anomalous quantum Hall effect. The edge and interface states of this system were also verified to be unidirectional and inmune to elastic backscattering. It was the first time that the topological magnon state was suggested in the skyrmion crystal, and the main results are shown in Fig. 5(b). Hence, many subsequent studies on 2D-SMCs focused on the topological characteristics in their magnon bands. In 2017, S. K. Kim *et al.* theoretically studied the magnon bands of the skyrmion lattice with honeycomb structure which is exhibited in Fig. 5(c)[93]. Their results obviously exhibited a chiral edge state in the magnon bands, which is shown by the blue and red arrows in the left side, as well as the blue and red lines in the right side of Fig. 5(c). Then in 2018, Z. X. Li *et al.* reported a coexistence of multiple chiral and non-chiral edge states of the magnon bands in honeycomb skyrmion lattices. They attributed the non-chiral edge states to the Tamm-Shockley mechanism, and attributed the chiral ones with different handedness to the second-order inertial term of skyrmion mass as well as a third-order non-Newtonian gyroscopic term in the massless Thiele's equation. The results are revealed in Fig. 5(d)[94]. More recently, S. A. Díaz *et al.* and T. Hirosawa *et al.* reported several investigations on the topological nontrivial magnon bands of the 2D antiferromagnetic skyrmion lattice and the antiskyrmion lattice, as well as the tunable magnon bands from topological nontrivial to trivial in 2D ferromagnetic skyrmion lattice, which will be introduced in the following[95,96,98]. These studies suggest great potential applications of topological nontrivial 2D-SMCs for the magnonic devices based on the chiral and low-dissipation edge states.

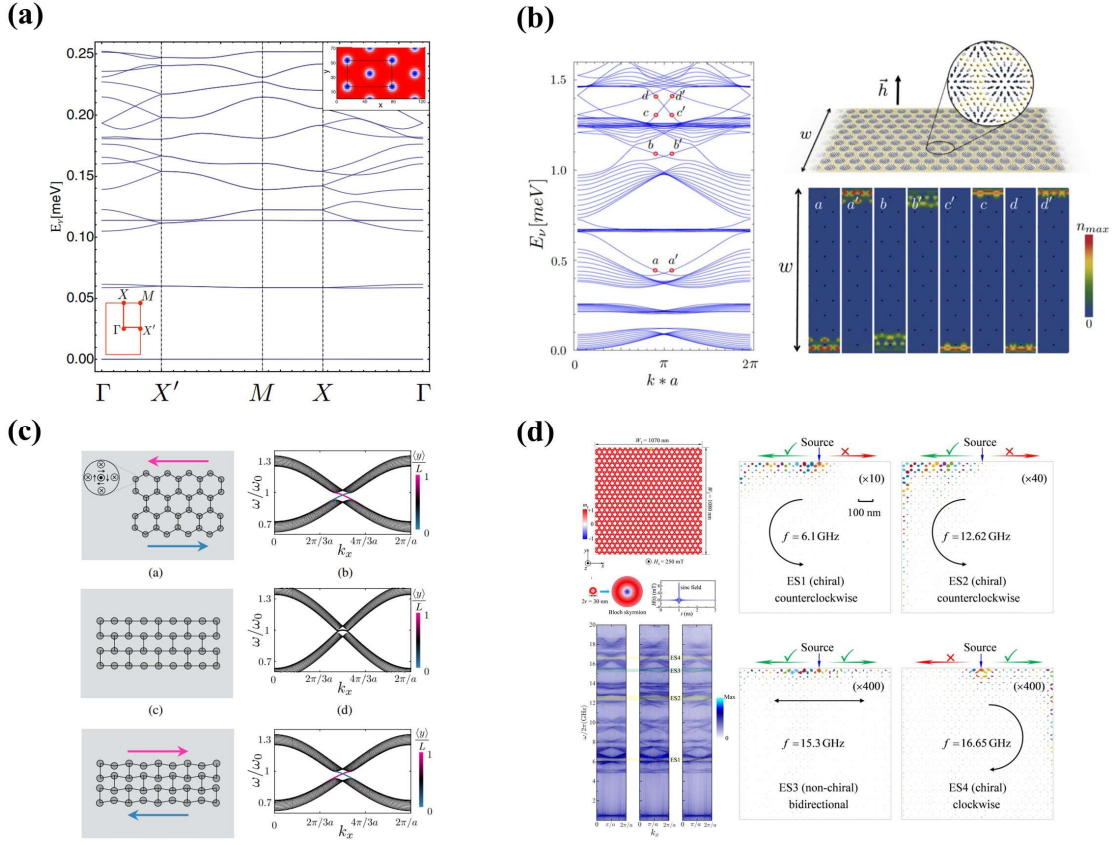

Fig. 5. (a) Magnon bands calculated for the 2D skyrmion crystal considering a two-skyrmions unit cell[104]. (b) The topological nontrivial magnon bands and the chiral edge states of 2D triangular SMCs based on Fe/Ir multilayers[92] (c) The geometric schematic illustration and the one-dimensional dispersion of skyrmion bubbles in a honeycomb lattice with zigzag edges, a rectangular lattice and a distorted rectangular lattice, respectively. [93]. (d) The sketch of a large-scale honeycomb skyrmion lattice, the magnon band structures and the real spatial profiles of the non-chiral and chiral edge states under different frequencies. [94]. [(a) is reproduced with the permission from A. Roldán-Molina, *et al.* Phys. Rev. B 92, 245436 (2015). Copyright 2015, American Physical Society. (b) is reproduced with the permission from A. Roldán-Molina, *et al.* "Topological spin waves in the atomic-scale magnetic skyrmion crystal", New J. Phys. Vol. 18, 045015, 2016, DOI: 10.1088/1367-2630/18/4/045015, licensed under a Creative Commons Attribution CC-BY 3.0 license. (c) is reproduced with the permission from S. K. Kim *et al.* Phys. Rev. Lett. 119, 077204 (2017). Copyright 2017, American Physical Society. (d) is reproduced with the permission from Z.-X. Li, *et al.* Phys. Rev. B 98, 180407(R) (2018). Copyright 2017, American Physical Society.]

## 4-3. Antiferromagnetic skyrmion based magnonic crystals and antiskyrmion magnonic crystals

Besides of the SMCs constituted by Néel- and Bloch-type skyrmions, there are other types of SMCs which are constituted by some general skyrmion structures, such as antiferromagnetic skyrmion based magnonic crystals (AFM-SMCs) and antiskyrmion magnonic crystals (ASMCs). AFM-SMCs are SMCs based on antiferromagnetic materials with DMI. In 2015, H. D. Rosales *et al.* predicted an antiferromagnetic skyrmion crystal (AFM-SkX) state in a pure antiferromagnetic frustrated system[172–174]. In 2019, S. A. Díaz *et al.* discovered that AFM-SkXs could also provide a possibility to obtain topologically protected magnonic edge states at sufficiently low energies, which suggested the AFM-SkXs to be a novel type of topological SMCs[95]. Fig. 6(a) exhibits the classical ground-state texture of the AFM-SkXs, the topological nontrivial magnon bands and the chiral edge states in momentum space and real space.

Antiskyrmions are firstly observed in acentric tetragonal Heusler compounds with $D_{2d}$ crystal symmetry by A. K. Nayak *et al.* in 2017[175]. Then in 2020, T. Hirosawa *et al.* declared that the SMCs constituted by antiskyrmions can provide higher-order topological phases of magnon modes[98]. The sketch of ASMCs and the higher-order topological magnon modes are shown in Fig. 6(b). This kind of higher-order topological magnon modes possess lower spatial dimension than the topological edge states and localized at the hinges or corners of the ASMCs. This research provides new approach to obtain higher-order topological phase in SMCs.

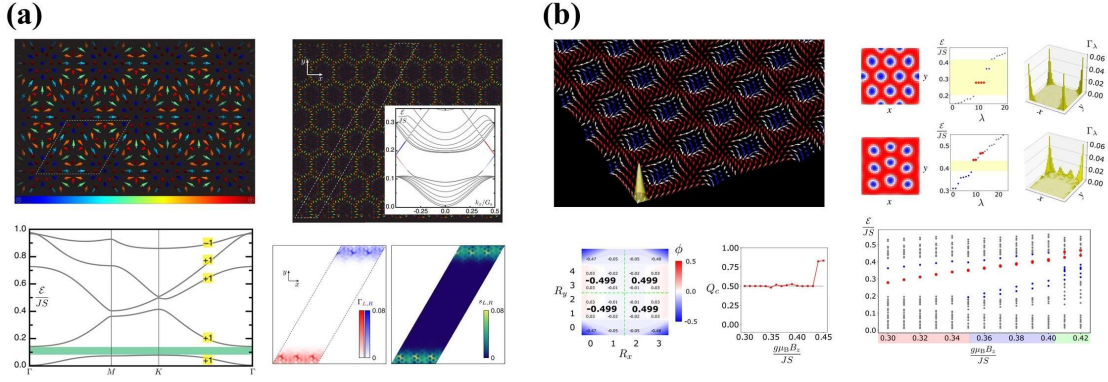

Fig. 6. (a) The classical ground-state texture of the AFM-SkX, the bulk magnon band structure and the Chern number of the bands, the AFM-SkX on a strip of infinite length along the x axis and its magnon band structure with the chiral edge states in the first band gap, and the chiral edge states in real space[95]. (b) The ground-state texture of the antiskyrmion crystals, the formation of high-order topological corner states of magnons in confined antiskyrmion crystals, and the Fractional corner charges induced by the quantized bulk quadrupole moment[98]. [(a) is reproduced with the permission from S. A. Díaz, *et al.* Phys. Rev. Lett. 122, 187203 (2019). Copyright 2019, American Physical Society. (b) is reproduced with the permission from T. Hirosawa, *et al.* Phys. Rev. Lett. 125, 207204 (2020). Copyright 2020, American Physical Society.]

**4-4. Experimental achievements of skyrmion based magnonic crystals.**

In past few years, there are plenty of theoretical and numerical works focused on the magnonic properties of SMCs. However, the experimental realization of SMCs is still an open issue[100]. As early as in 2010 and 2011, X. Z. Yu *et al*. reported a real space observation of a two-dimensional skyrmion crystal in $Fe_{0.5}Co_{0.5}Si$ and Fe/Ir at low temperature[145,171]. In 2012, Y. Onose *et al*. experimentally studied the magnetic excitations of natural skyrmion lattice in a helimagnetic insulator $Cu_2OSeO_3$ and observed two elementary excitations of the Skyrmion: the counterclockwise circulating mode at 1 GHz with the magnetic field polarization parallel to the Skyrmion plane and the breathing mode at 1.5 GHz with a perpendicular magnetic field polarization[161], the results is given in Fig. 7(a). And subsequently several novel experimental studies were reported which focused on the magnetic dynamic properties of these natural skyrmion systems[101–103]. In 2015, T. Schwarze *et al*. quantitatively studied the collective dynamic processes of the entire magnetic phase (include the skyrmion lattice phase) in MnSi, $Fe_{1-x}Co_xSi$ and $Cu_2OSeO_3$, and obtained the magnetic parameters from the dynamic experiment data, as shown in Fig. 7(b) [101]. In 2020, S. Seki *et al.* reported the propagation of the skyrmion dynamic excitations along the skyrmion strings in $Cu_2OSeO_3$. They found that this propagation is directionally non-reciprocal and the degree of non-reciprocity, as well as group velocity and decay length, are strongly dependent on the character of the excitation modes (see Fig. 7(c))[102]. Then in 2021, A. Aqeel *et al.* reported new resonant modes in $Cu_2OSeO_3$ [103], the main results are shown in Fig. 7(d). They identified resonant modes associated with the tilted conical state, the gyrational and breathing modes associated with the low temperature skyrmion state, as well as the hybridization of the breathing mode with a dark octupole gyration mode. Especially, the resonances observed in small fields after field cycling indicate the presence of elongated skyrmions, which suggest that under decreasing fields the hexagonal skyrmion lattice becomes unstable with respect to an oblique deformation. However, although there are several works focus on the dynamic modes in natural skyrmion lattices in chiral magnets such as MnSi, $Fe_{0.5}Co_{0.5}Si$ and $Cu_2OSeO_3$, the magnon band structures of these skyrmion lattices still need further investigations. Besides, the presence of the natural skyrmion lattice in these magnets rely on low temperature, which limits their potential in magnonic applications.

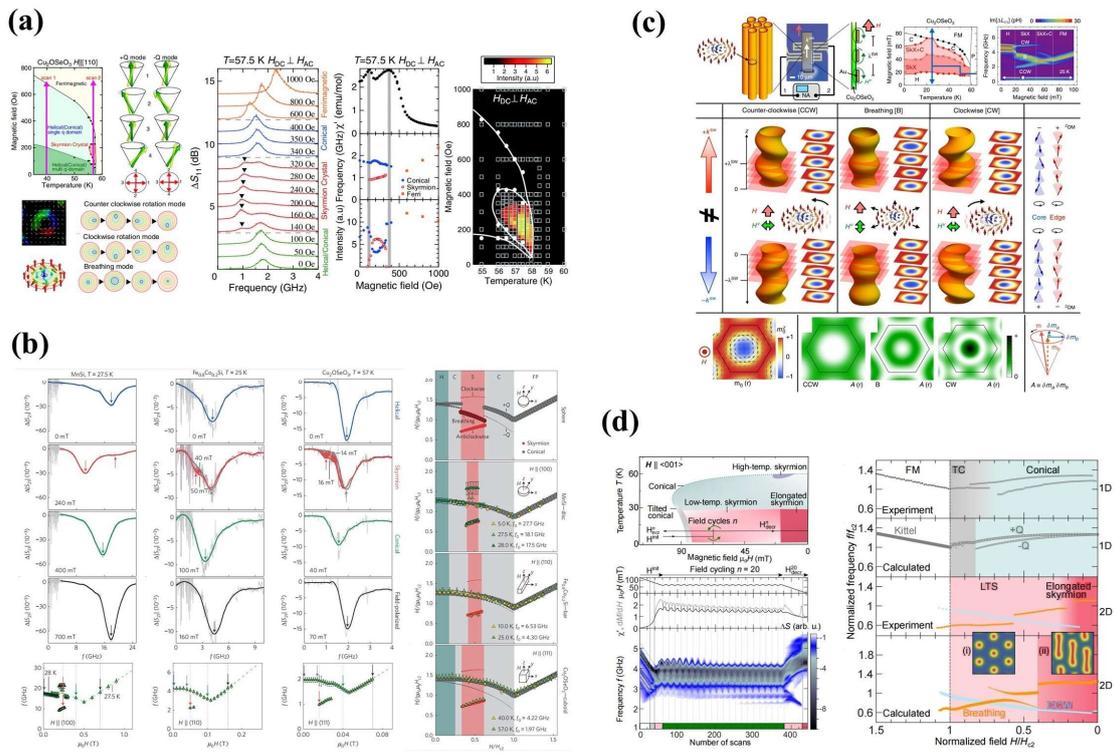

Fig. 7. (a) The counterclockwise circulating mode and the breathing mode in the skyrmion lattice of $Cu_2OSeO_3$ [161]. (b) The collective excitation of skyrmion lattices in MnSi, $Fe_{1-x}Co_xSi$ and $Cu_2OSeO_3$ [101]. (c) The skyrmion excitation propagating along the skyrmion strings[102]. (d) The tilted conical state, conical state, skyrmion gyrational and breathing modes in $Cu_2OSeO_3$, as well as the illustration of the skyrmion lattice and the elongated skyrmion lattice[103]. [(a) is reproduced with the permission from Y. Onose, *et al.* Phys. Rev. Lett. 109, 037603 (2012). Copyright 2012, American Physical Society. (b) is reproduced with the permission from M. Garst, *et al.* "Collective spin excitations of helices and magnetic skyrmions: review and perspectives of magnonics in non-centrosymmetric magnets", J. Phys. D: Appl. Phys. Vol. 50, 293002, 2017, DOI: 10.1088/1361-6463/aa7573; licensed under a Creative Commons Attribution CC-BY 3.0 license. (c) is reproduced with the permission from S. Seki, *et al.* "Propagation dynamics of spin excitations along skyrmion strings", Nat. Commun. Vol. 11, 256, 2020; licensed under a Creative Commons Attribution 4.0 International License. (d) is reproduced with the permission from A. Aqeel, *et al.* Phys. Rev. Lett. 126, 017202 (2021). Copyright 2021, American Physical Society.]

In order to overcome the shortage of the natural skyrmion systems and achieve the experimental characterizing of the magnon band structure of skyrmion lattice, the thin-film skyrmion systems came into people's view because they can provide stable skyrmions at room temperature. Unfortunately, it is difficult to arrange the spontaneous DMI-induced skyrmions into organized periodic arrays in the films. Thus, researchers developed several methods to obtain artificial skyrmion lattices in thin films without DMI. In the early 2010s, one route to realize artificial skyrmion lattices was reported which imprinted the magnetic vortexes of the top nanodots into the underlayer film with perpendicular magnetic anisotropy[162–164,166], as shown in Figs. 8(a) and 8(b). In 2020, Y. Li *et al.* provided a new method to build artificial skyrmion lattice in top-layer-nanostructured synthetic antiferromagnetic (SAF) multilayers via the interlayer antiferromagnetic coupling[165], which is illustrated in Fig. 8(c). In the same year, Y. Guang *et al.* reported another method to build zero-field skyrmion lattice at room temperature through soft X-ray irradiation in an exchange-biased [Pt/Co/IrMn]$_n$ magnetic multilayer[176]. They demonstrated that the soft X-ray can induce an unexpected exchange bias reorientation effect. Based on this effect, single skyrmions, skyrmion-track, and artificial skyrmion lattice were successfully created, as shown in Fig. 8(d). Nowadays, although the methods to build artificial skyrmion lattices are already developed, rare experimental study focus on the magnonic characteristics of artificial SMCs. There are two shortages in artificial skyrmion lattices which limit their application for SMCs: on the one hand, the artificial skyrmion lattices can exist only in an external field with small range, thus it is difficult to observe the dynamic properties of the skyrmion lattices clearly; on the other hand, the artificial skyrmion lattices usually based on the magnetic materials with high Gilbert damping, it limits the propagation of the SWs in these skyrmion lattices. According to these shortages, the key points to build artificial SMCs are enhancing the stability of the skyrmion lattices in external fields, and reducing the Gilbert damping of the films, both of which need further study.

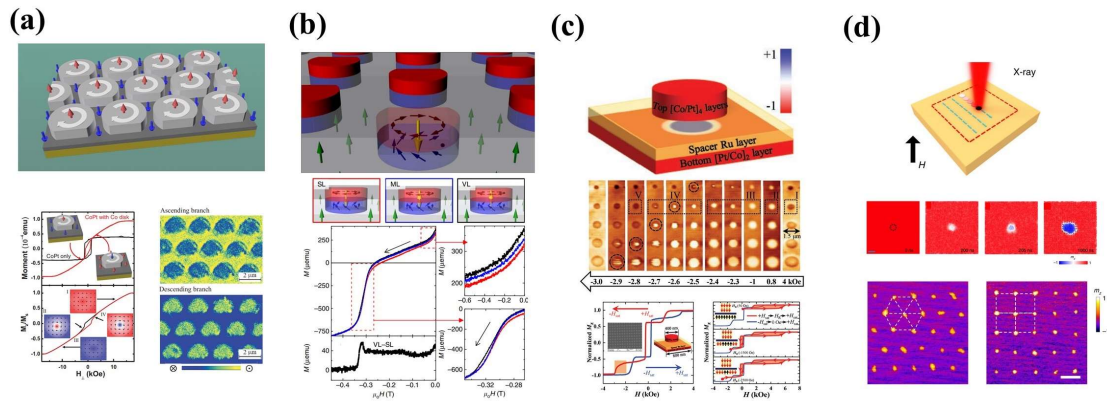

Fig. 8. (a)-(b) 2D-skyrmion crystals induced by ordered arrays of magnetic submicron disks prepared on top of films with perpendicular anisotropy, as well as their different magnetizing states under various of magnetic fields revealed by the hysteresis loops, polar Kerr images and magnetization curves[162–164]. (c) Diagram of artificial skyrmion configuration in [Co/Pt]$_4$/Ru/[Co/Pt]$_2$ SAF structure. And the measured room temperature MFM images and hysteresis loops of the nanostructured SAF multilayer[165]. (d) Sketch of using synchrotron X-rays to introduce a uniform exchange bias through scanning a closed area under a perpendicular magnetic field in exchange-biased [Pt/Co/IrMn]$_n$ magnetic multilayer. The snapshots of the single skyrmion creation process. And the artificial skyrmion lattice patterned by X-ray illumination[176]. [(a) is reproduced with the permission from B. F. Miao, *et al.* Phys. Rev. B 90, 174411 (2014). Copyright 2014, American Physical Society. (b) is reproduced with the permission from D. A. Gilbert, *et al.* "Realization of ground-state artificial skyrmion lattices at room temperature", Nat. Commun. Vol. 6, 8462, 2015. licensed under a Creative Commons Attribution 4.0 International License. (c) is reproduced with the permission from Y. Li, *et al.* Adv. Funct. Mater. 30, 1907140 (2020). Copyright 2020, Wiley Publishing (d) is reproduced with the permission from Y. Guang, *et al.* "Creating zero-field skyrmions in exchange-biased multilayers through X-ray illumination", Nat. Commun. Vol. 11, 949, 2020; licensed under a Creative Commons Attribution 4.0 International License.]

# 5. Perspective

**5-1. Topological phases and their evolution**

In recent years, a new subfield of magnonics called topological magnonics is developing rapidly. Research in this field is focused on the novel characteristics of topological nontrivial magnon modes[58,60,64,125,177,178]. In the view of theory, topological magnonics can provide a new access to help people understanding the topological nontrivial phenomena of magnonic systems in condensed matter, such as Dirac magnons, Weyl magnons and higher-order topological magnons[92,179–185]. In the view of application, topological magnonic systems can provide topological nontrivial magnon edge states or surface states with forbiddenness of backscattering, which can be applied in future magnonic devices with high speed and low energy consumption[186,187]. Topological MCs are ideal platforms for the studies of topological magnonics, which can provide topological nontrivial modes in their magnon bands. As previously mentioned, in 2013, R. Shindou *et al.* firstly reported a project of topological MCs with chiral edge mode[121]. And in 2016, A. Roldán-Molina *et al.* showed a new possibility to achieve topological nontrivial magnon bands and chiral edge states in SMCs[92]. Since then, much research is focused on the topological magnon modes in SMCs[93–98].

Because the magnon band structures and band gaps of SMCs can be switched dynamically, the modulation and transition of the magnon modes in SMCs from topological nontrivial to trivial become new issues to discuss. In 2020, X. Wang *et al.* provided a topological transition study of the magnon band structure in artificial skyrmion lattice induced by nanopatterned vortexes. In this work they use external field to control the existence of the artificial skyrmion lattice, and thus switch the chiral edge state from "on" to "off", as shown in Fig. 9(a)[97]. In the same year, S. A. Díaz *et al.* introduced a controllable topological switching between the nontrivial phase and trivial phase in SMCs without destroying the skyrmion lattice via tunning external magnetic fields, which is shown in Fig. 9(b)[96]. In this work, they discovered that in specific triangular SMCs, while setting the external field from lower to higher than a critical field, the topological nontrivial magnon band gap can be transformed to topological trivial, simultaneously the chiral edge states will vanish. These works provided a new topic on modulating the topological phase of SMCs by external effects for further tunable topological magnon devices.

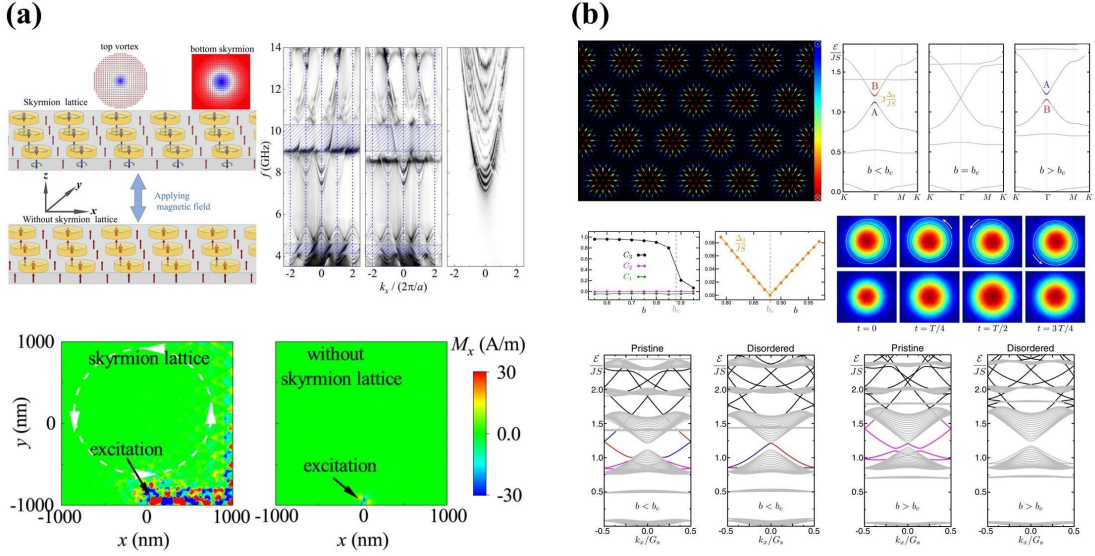

Fig. 9. (a) The sketch of the artificial skyrmion lattice induced by nanopatterned magnetic vortexes, and the magnon bands of the honeycomb artificial skyrmion lattice with chiral edge states, as well as the real spatial illustration of the edge states, respectively. It is pointed out that the chiral edge states would vanish while the skyrmion lattice is annihilated by the external magnetic field[97]. (b) The sketch of the skyrmion lattice, as well as the evolution of the magnon band structure, Chern number of the three lowest energy magnon bands, band gap width, and the chiral edge states with the changing external field. It can be pointed out that while the external field change from lower to higher than the critical field $b_c$, the topological nontrivial band gap switches to trivial and the chiral edge states vanish. [96] [(a) is reproduced from X. Wang, *et al.* J. Appl. Phys. 128, 063901 (2020). with the permission of AIP Publishing. (b) is reproduced with the permission from S. A. Díaz, *et al.* "Chiral magnonic edge states in ferromagnetic skyrmion crystals controlled by magnetic fields", Phys. Rev. Res. Vol. 2, 013231, 2020, DOI: 10.1103/PhysRevResearch.2.013231, licensed under a Creative Commons Attribution 4.0 International License.]

## 5-2. Hybrid and quantum phenomena of magnons

Recently, hybrid magnonics have attracted much attention as a new subfield of magnonics[65,67,188–190]. The research on hybrid magnonics focuses on the coupling effects between magnons and other quantum systems such as microwave photons[67,191–200], optical photons[188,201–204], phonons[196,205–212] or magnons themselves[71,213–217]. These hybrid phenomena show a new possibility to apply magnons for quantum information field[65,188,189,218]. However, most of the research on hybrid phenomena of magnons are based on uniform magnetic media. Only a few works concerned the role of SMCs in these phenomena very recently[103,105]. As mentioned in Section 4-4, A. Aqeel *et al.* firstly reported an investigation on the hybrid of magnon modes in the low temperature skyrmion lattice state of $Cu_2OSeO_3$ in 2021[103]. In this work, they found that there is a hybridization of the breathing mode and a dark higher-order counterclockwise gyration mode, which performs as an anticrossing in the breathing mode of the skyrmion lattice. Besides, this hybridization can be modulated by the magnetocrystalline anisotropy $K$ (see Fig. 10(a)). In the same year, L. Liensberger *et al.* experimentally studied the cooperativity of magnon–photon coupling in $Cu_2OSeO_3$ with different magnetic phases, including the high-temperature skyrmion lattice state[105]. In this work, they modulated the magnetic phases of $Cu_2OSeO_3$ with tunning external magnetic field, and observed a strong tunability of the normalized coupling rate and the magnon-photon cooperativity by magnetic field at the phase boundaries of the skyrmion lattice state. The results are shown in Fig. 10(b). These novel investigations provide a new platform of skyrmion lattice for hybrid magnonics.

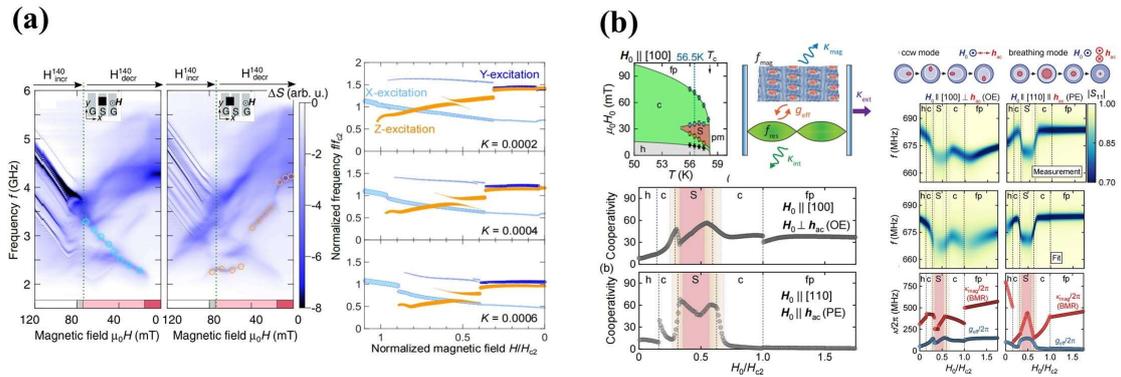

Fig. 10. (a) The experimental and theoretical results of the dynamic modes of the low temperature skyrmion lattice in $Cu_2OSeO_3$[103]. The breathing mode is divided into three parts by an anticrossing and a break. The anticrossing is generated by the hybridization of the breathing mode and a dark higher-order counterclockwise gyration mode. (b) The tunable cooperativity of magnon-photon system in $Cu_2OSeO_3$ modulated by external fields. The magnon-photon cooperativity is dramatically tuned by external fields at the phase boundaries of the skyrmion lattice state[105]. [(a) is reproduced with the permission from A. Aqeel, *et al.* Phys. Rev. Lett. 126, 017202 (2021). Copyright 2021, American Physical Society. (b) is reproduced with the permission from L. Liensberger, *et al.*

While consider the coupling effects between magnons and other (quansi-) particle systems in quantum limit (consider the coupling between single magnon and other quanta), hybrid magnonics will be linked up to quantum magnonics. Quantum magnonics is another new subfield of magnonics which is aimed to clarify the quantum characteristics of magnons (such as Bose-Einstein condensation and magnon Josephson effect), and try to connect magnons to qubits in quantum information devices via the coupling mentioned above[66,188,189,218–221]. However, the study of quantum magnonic phenomena in SMCs is still lacking. In 2020, A. Mook *et al.* reported a new theoretical study on a quantum-mechanical many-body phenomenon of propagating magnons in SMCs which is called spontaneous quasiparticle decay (SQD) [106]. SQD of magnons is a nonlinear magnon scattering process during which a propagating magnon is scattered into two other magnons while the initial magnon mode is close to the flat magnon bands corresponding to the anticlockwise, breathing, or clockwise mode of the skyrmion lattice. But till now, the quantized coupling between magnons and other (quansi-) particle systems is an open issue which needs more investigations.

## 6. Conclusion

In this review we provide a brief review on the concept of SMCs and the study situation on their interesting characteristics in magnonics. Firstly, we introduce the concepts of MCs and magnetic skyrmions. Then we show a full view on the theoretical and numerical progresses on the magnon band characteristics of SMCs comprised by ferromagnetic skyrmions, antiferromagnetic skyrmions, and antiskyrmions. In view of these theoretical and numerical studies, the magnon bands of SMCs exhibit significant characteristics, such as dynamic tunability and topological nontrivial magnon modes. Besides, we also present the experimental investigations of the dynamic modes in skyrmion lattices, and the fabrication of the artificial skyrmion lattices with flexible spatial periodicity. At the end, we give an outlook and perspectives on new fascinating topics based on SMCs, such as topological magnonics, hybrid magnonics, and quantum magnonics. SMCs show great potential value in these subfields of magnonics, which are declared to possess topological transition, hybrid phenomena between different dynamic modes, and quantum damping induced by SQD.

## ACKNOWLEDGEMENTS

This work was supported by the National Natural Science Foundation of China (Grant Nos. 12074189 and 11704191).

## DATA AVAILABILITY

The data that support the findings of this study are available from the corresponding author upon reasonable request.

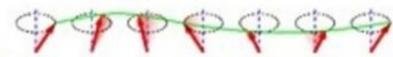

# Magnonics

**Physics**  **Application**

**Topological magnonics**

**Hybrid magnonics**  **Logic gates**

**Magnon transistors**

**Quantum magnonics**  **Interferometers**

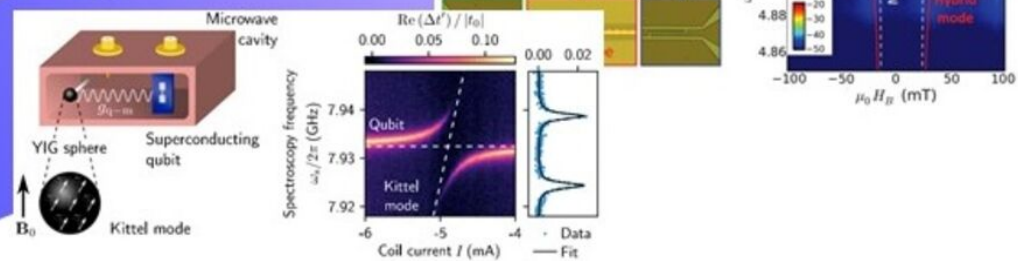
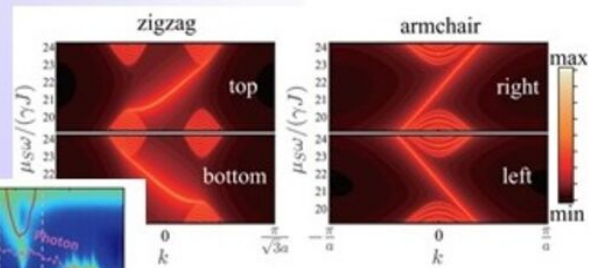
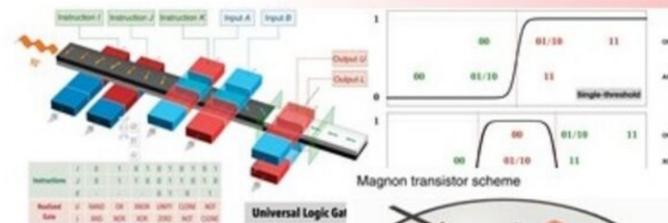
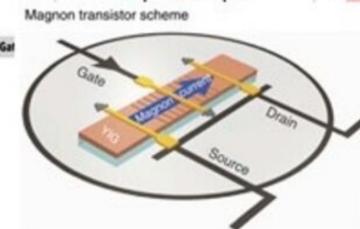
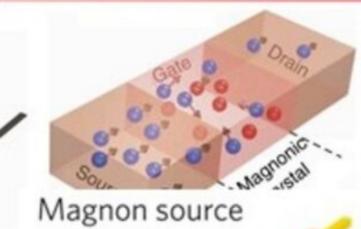
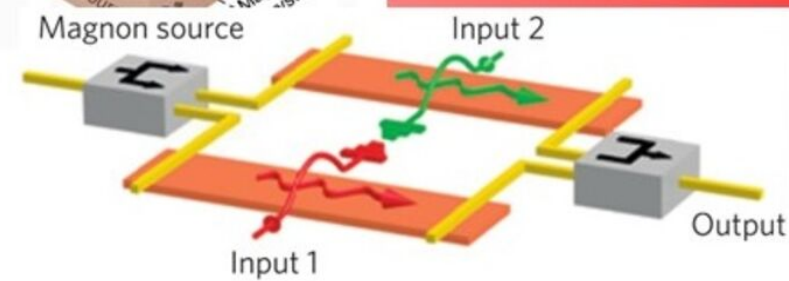

## (a) 1D-MCs

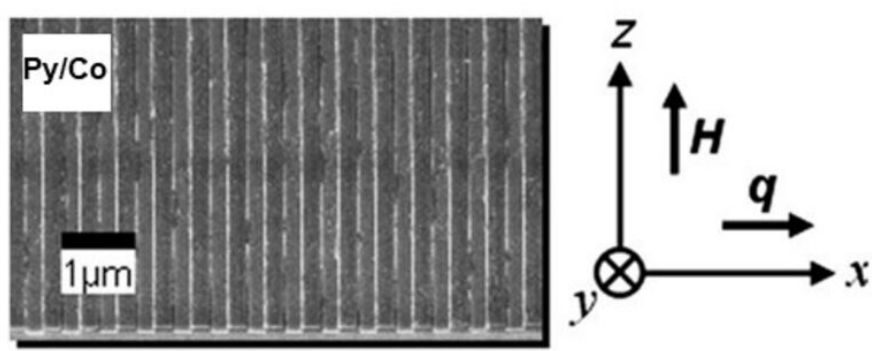
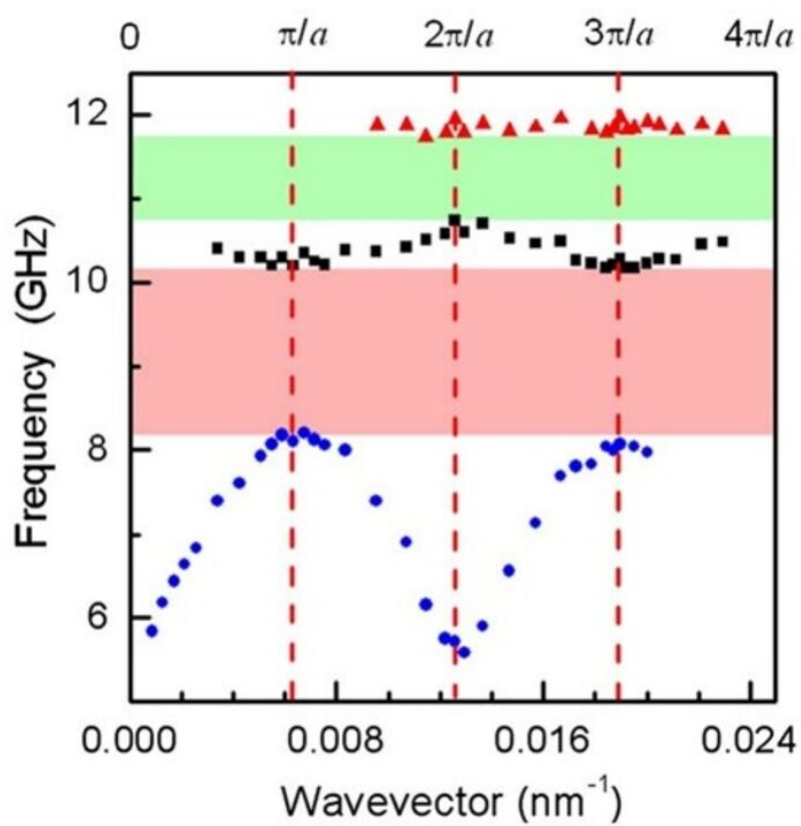

## (b) 2D-MCs

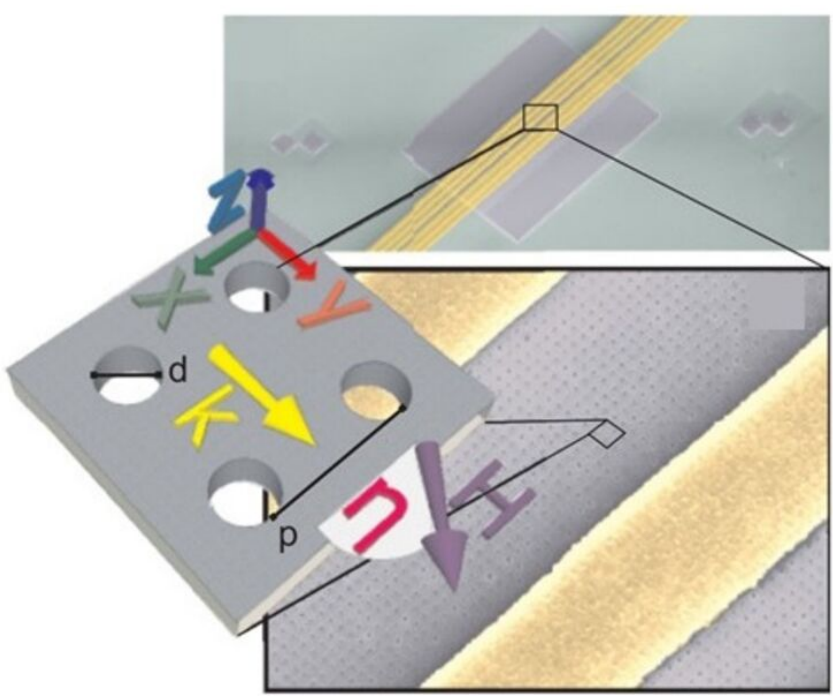
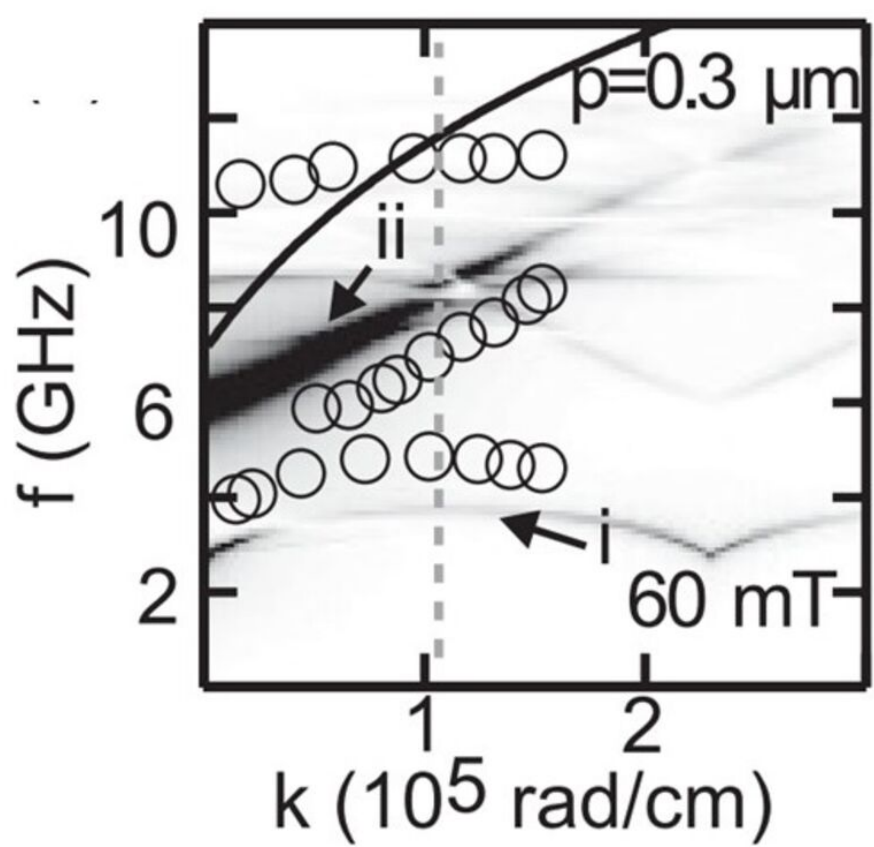

## (c) 3D-MCs

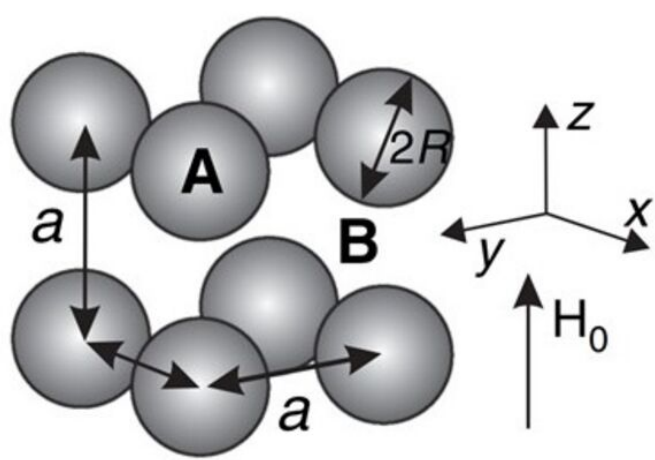
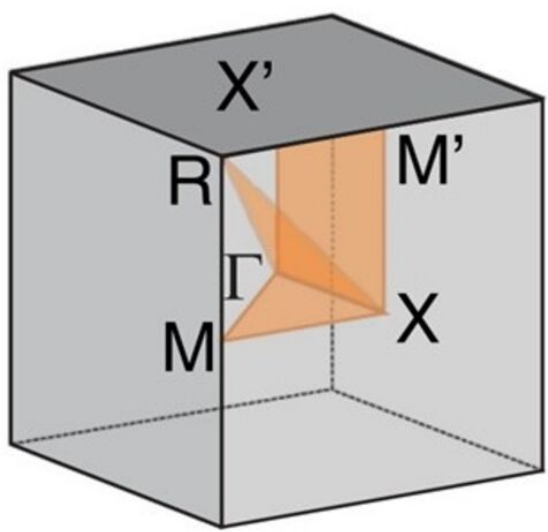
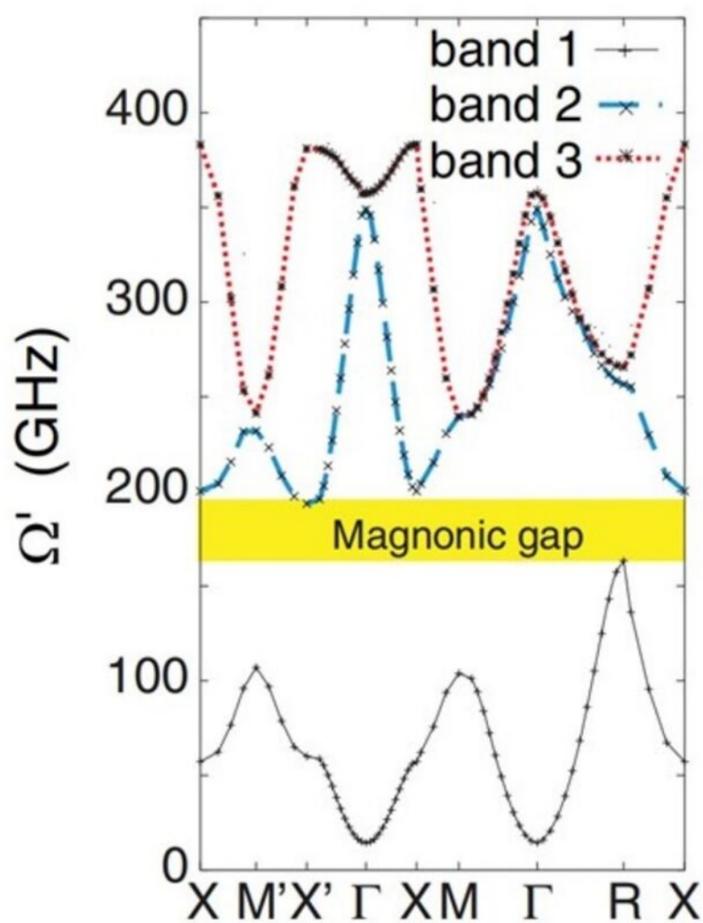

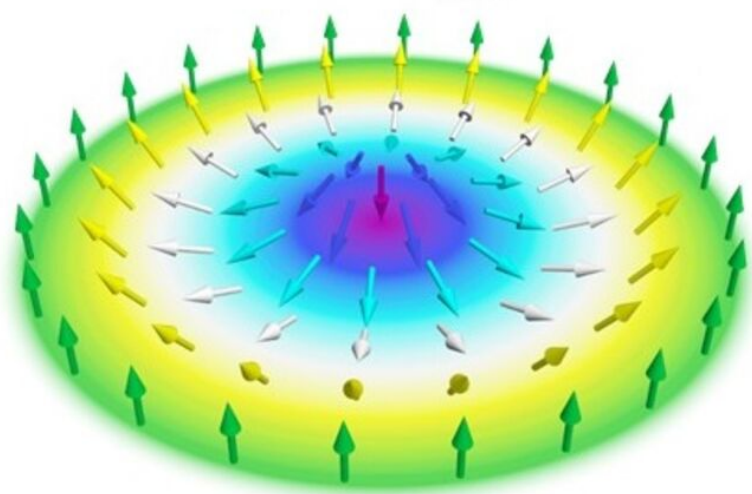
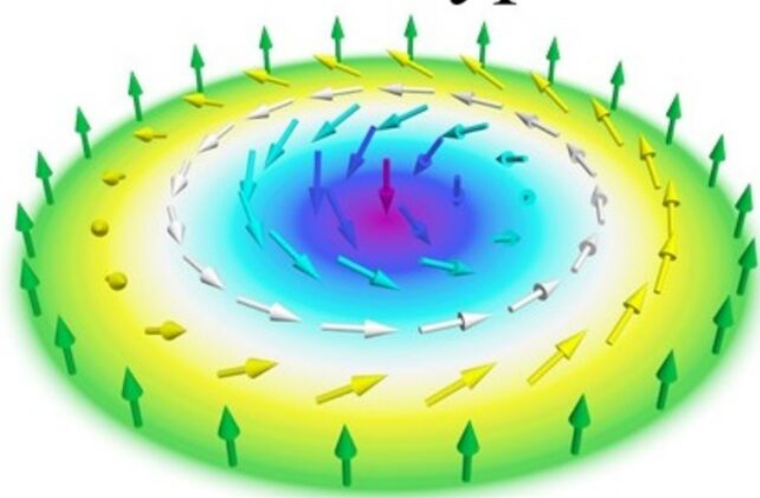
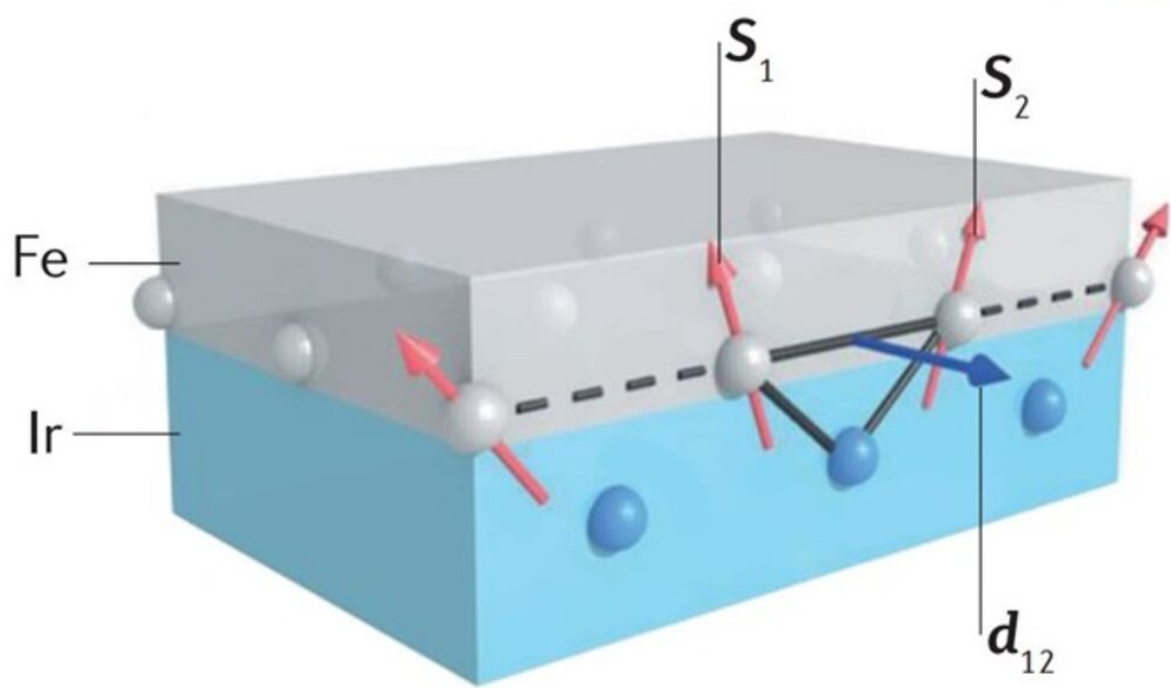
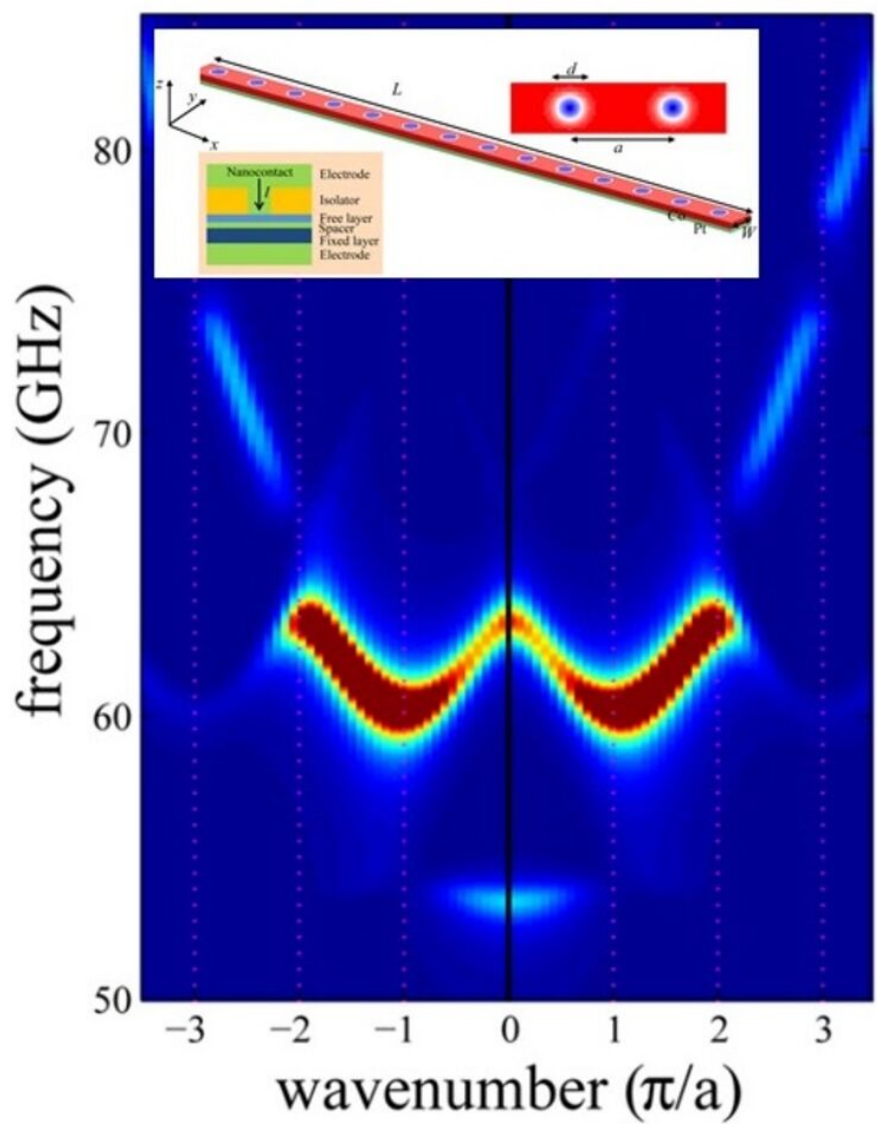
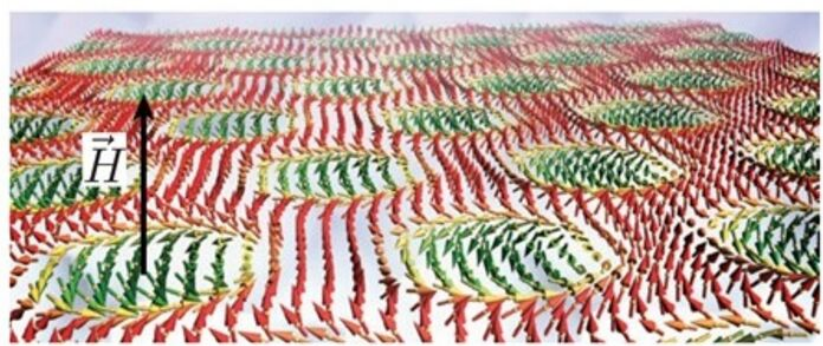
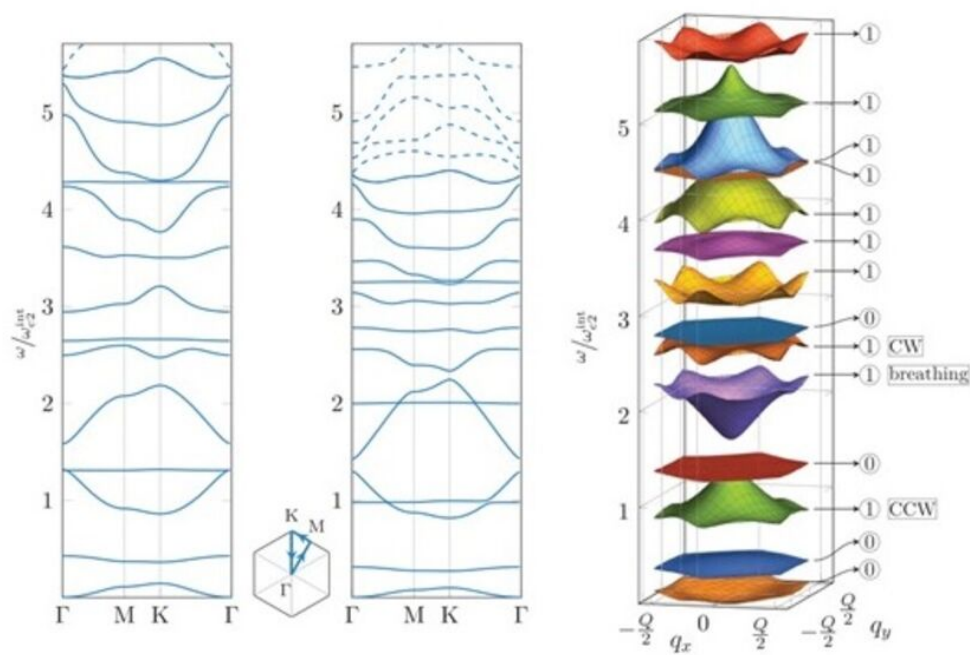

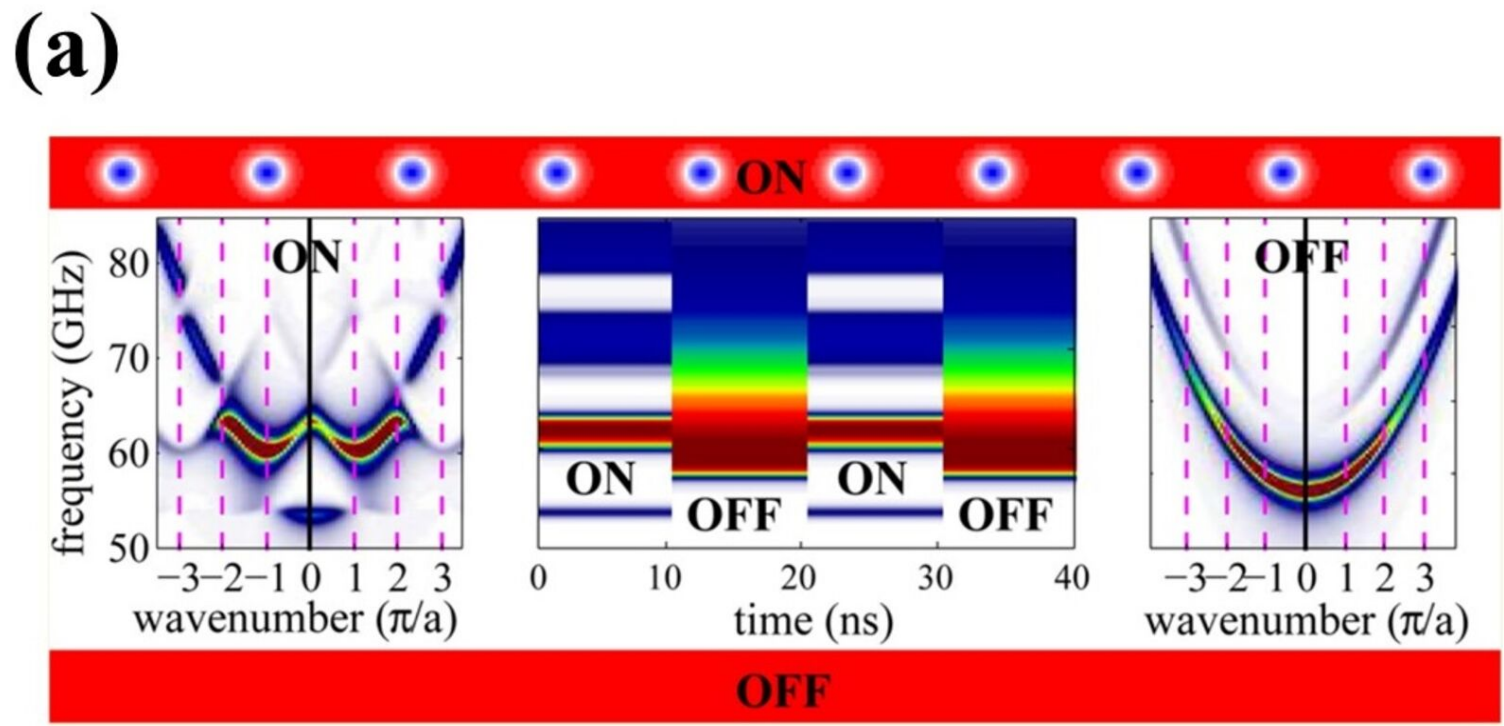
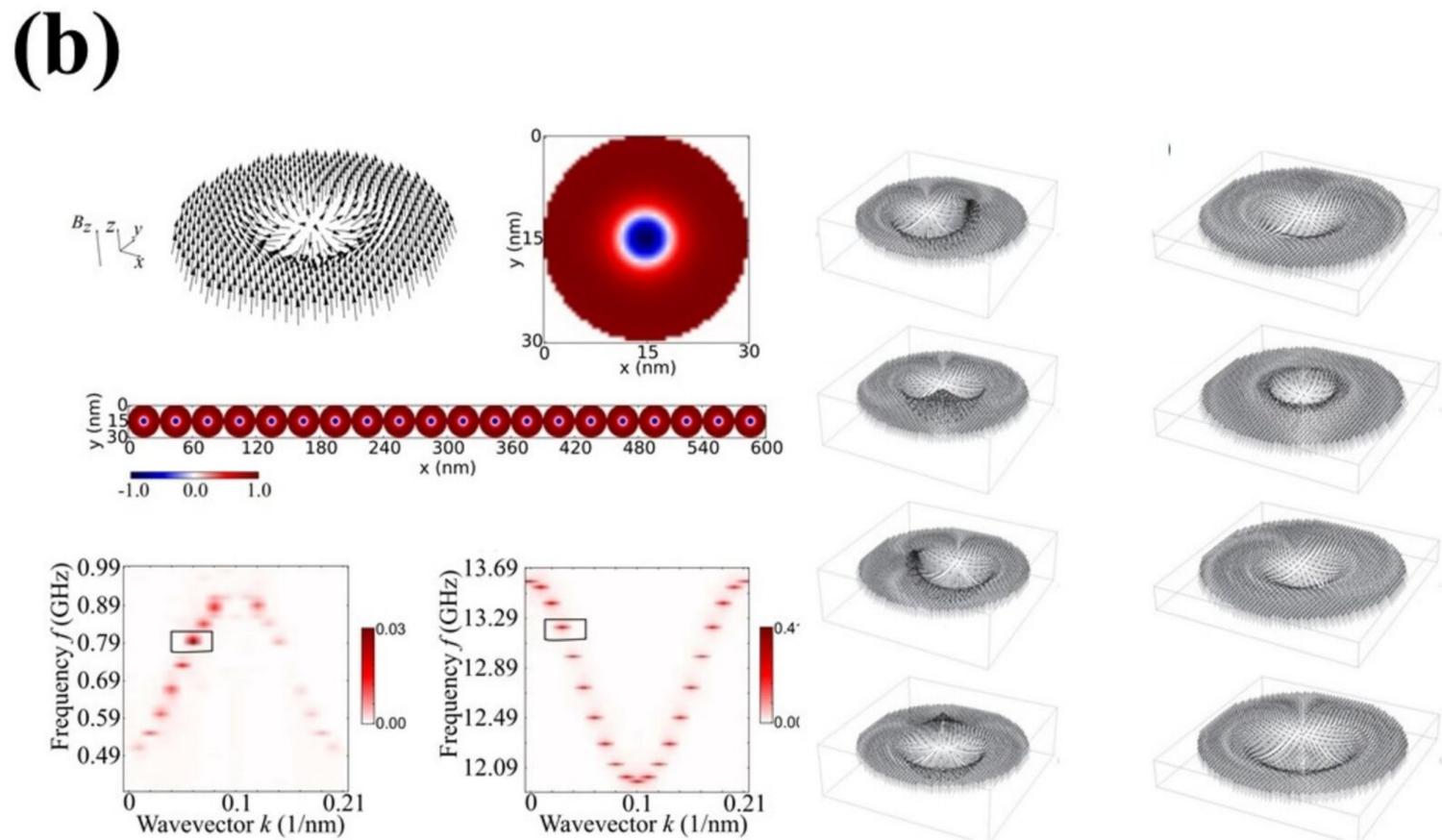
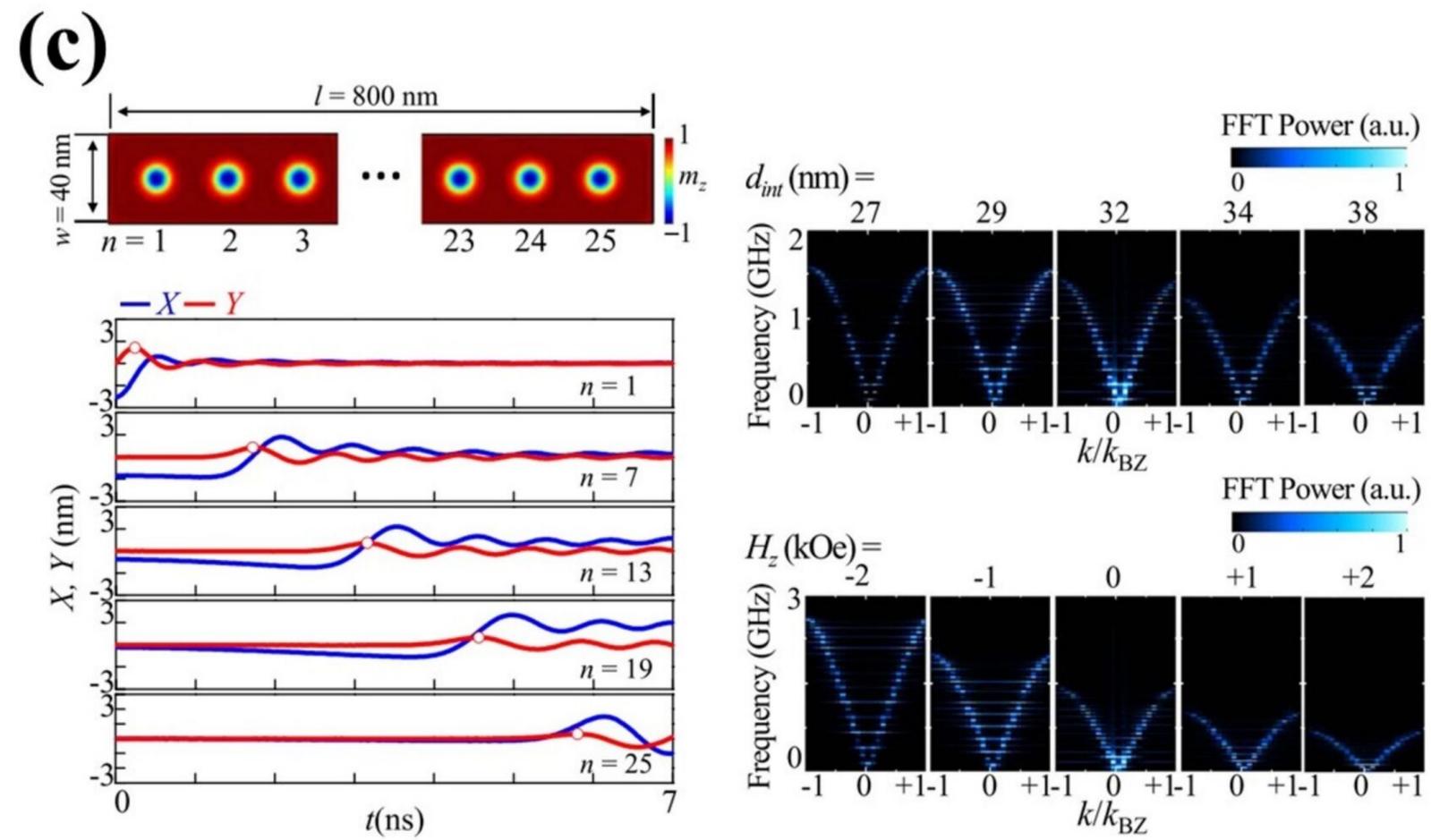
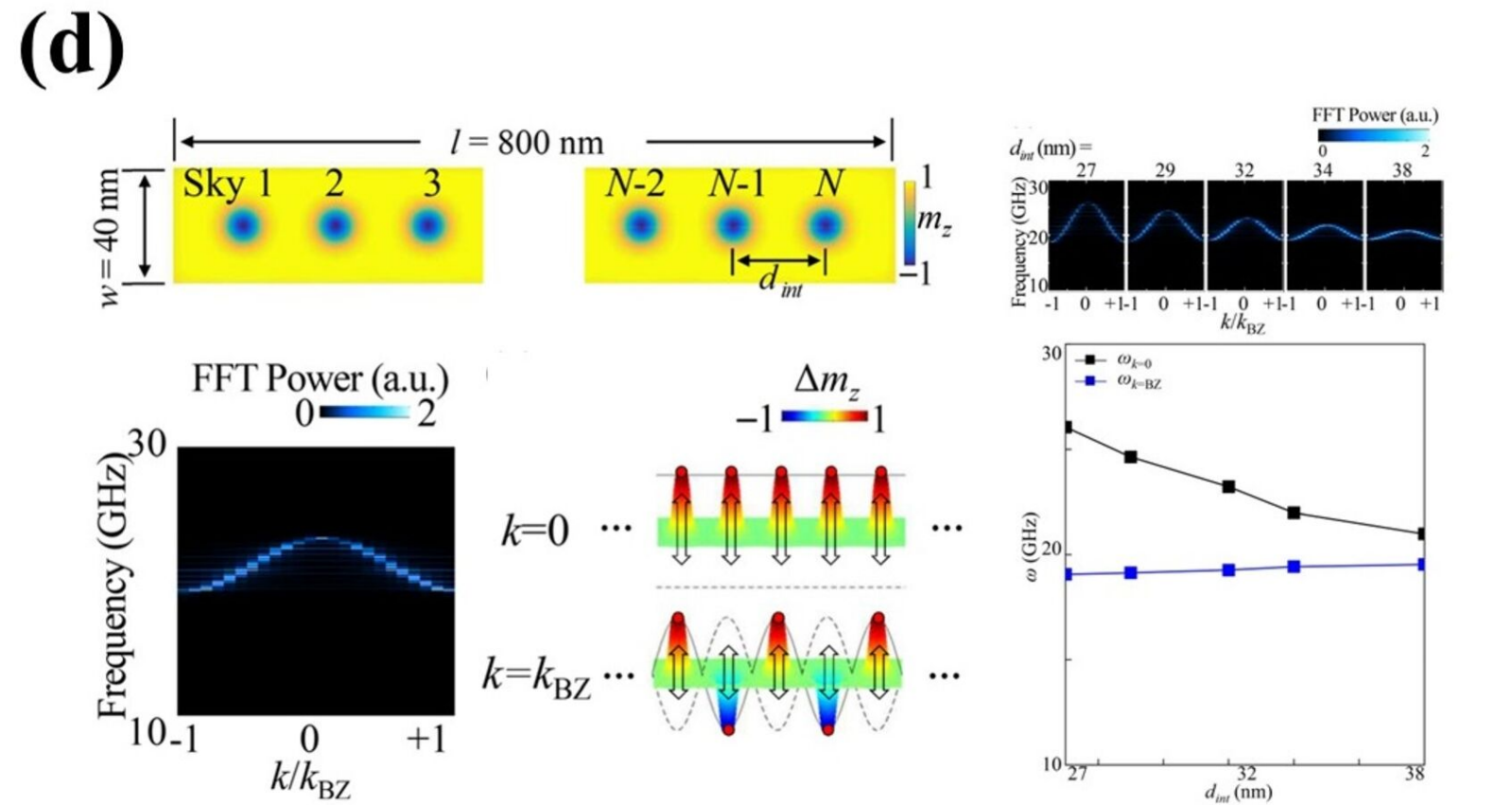

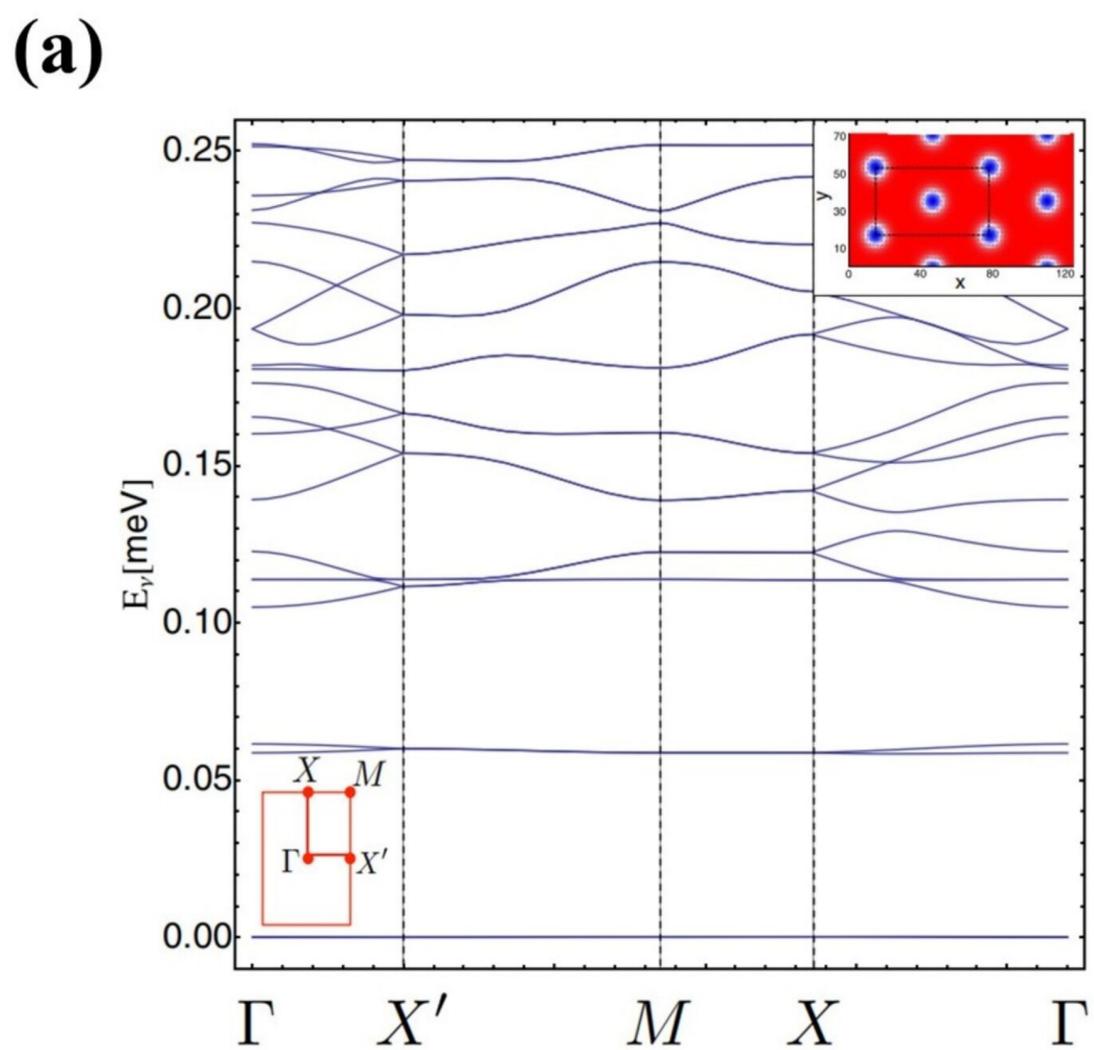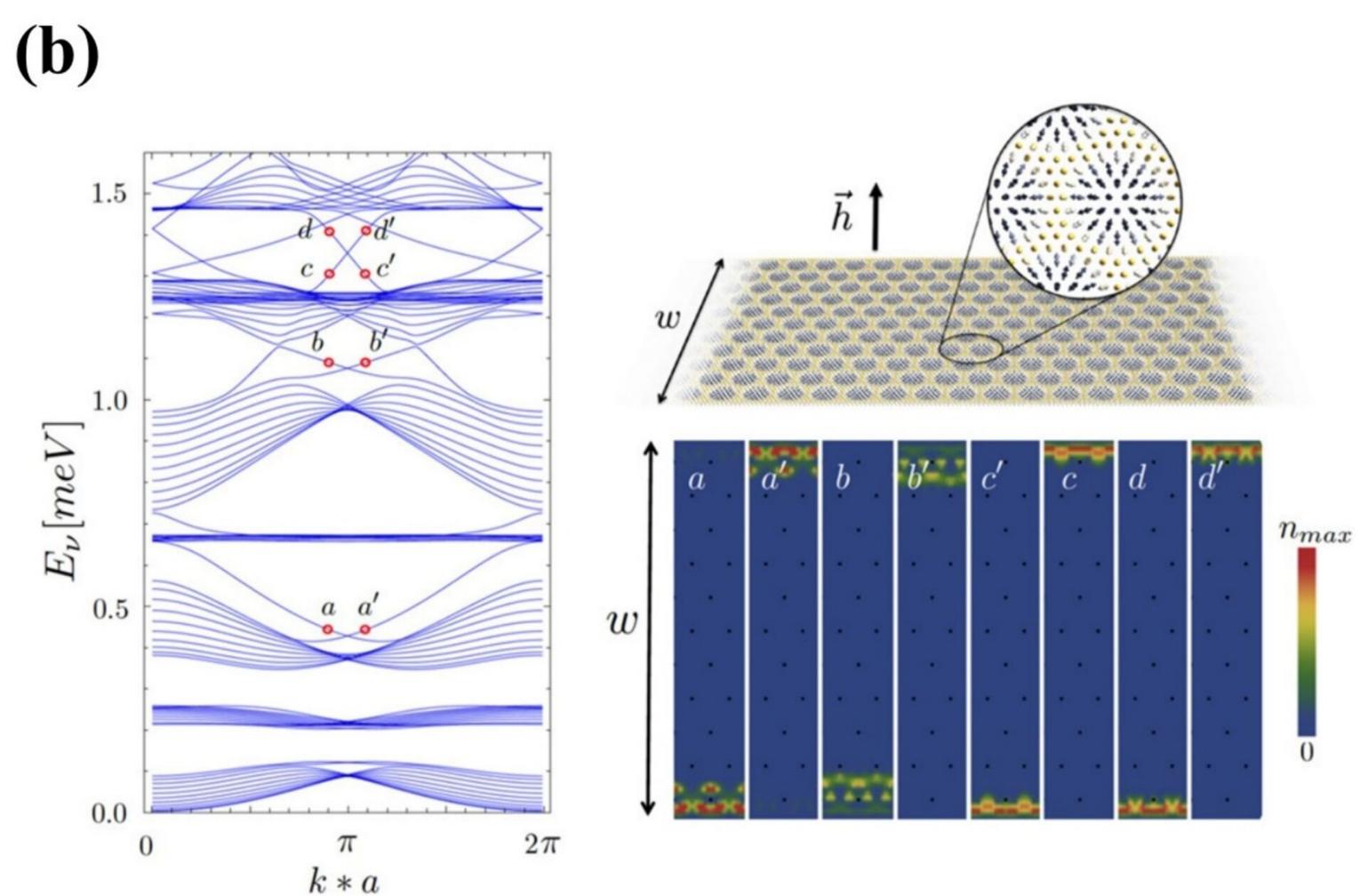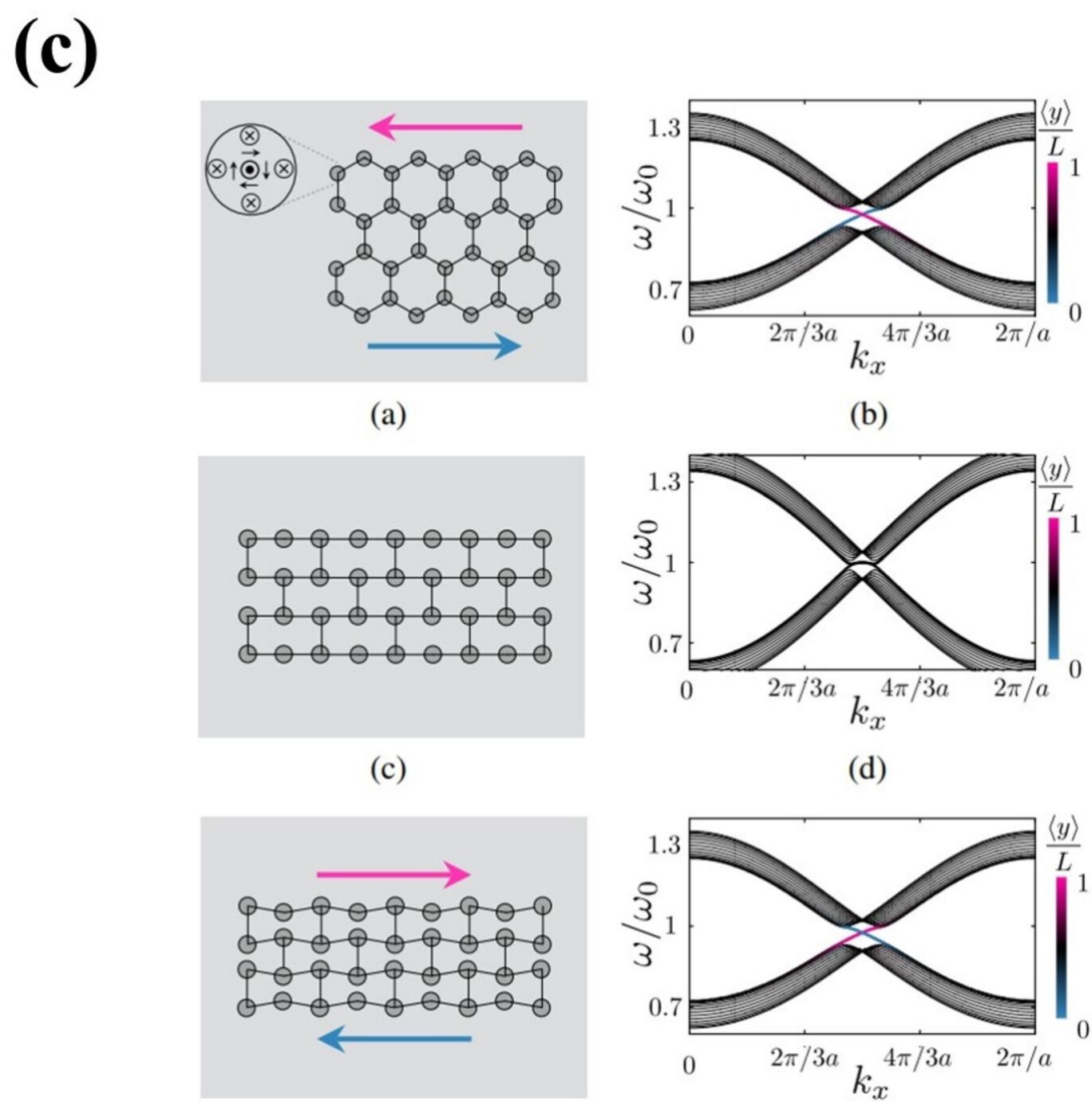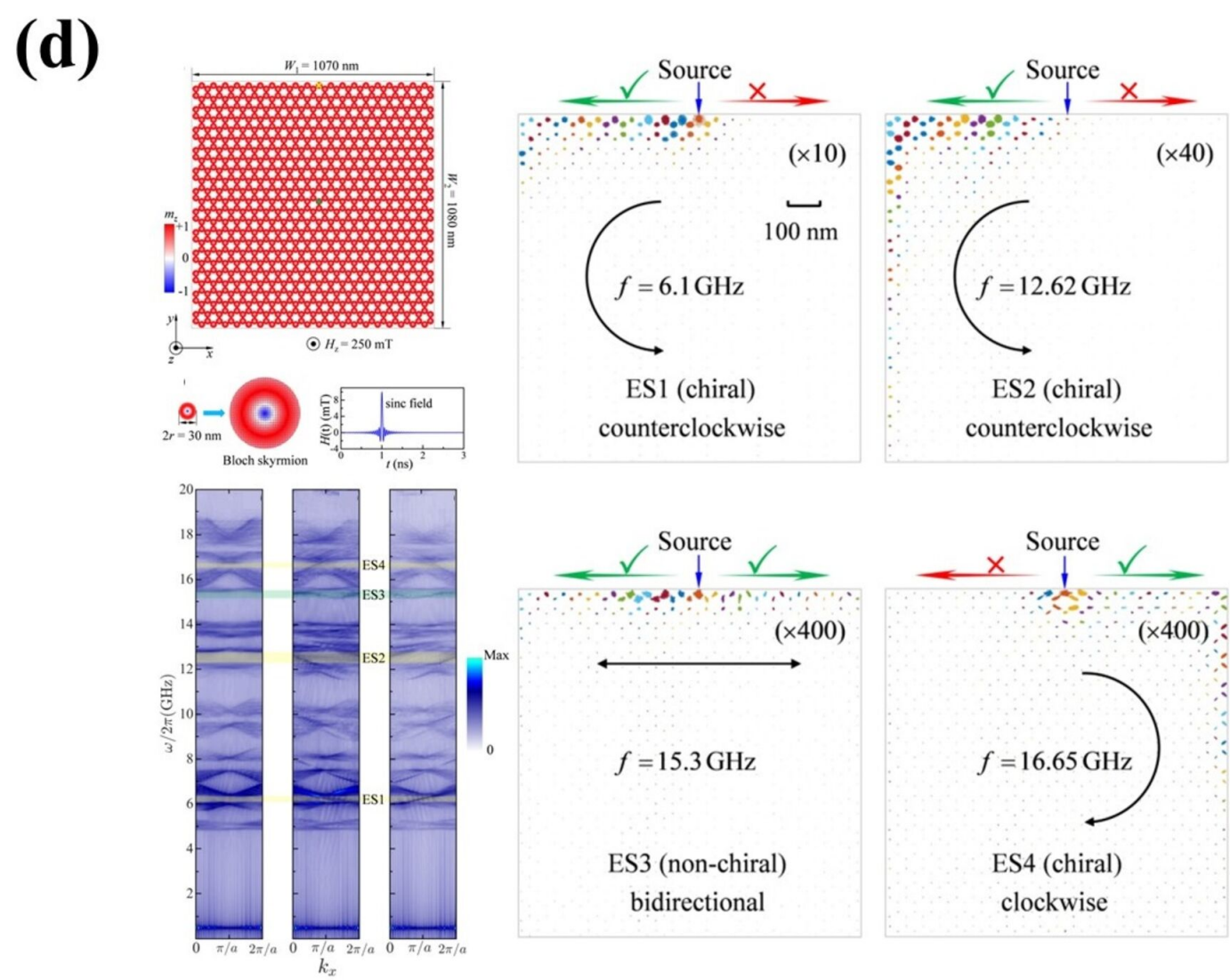

**(a)** 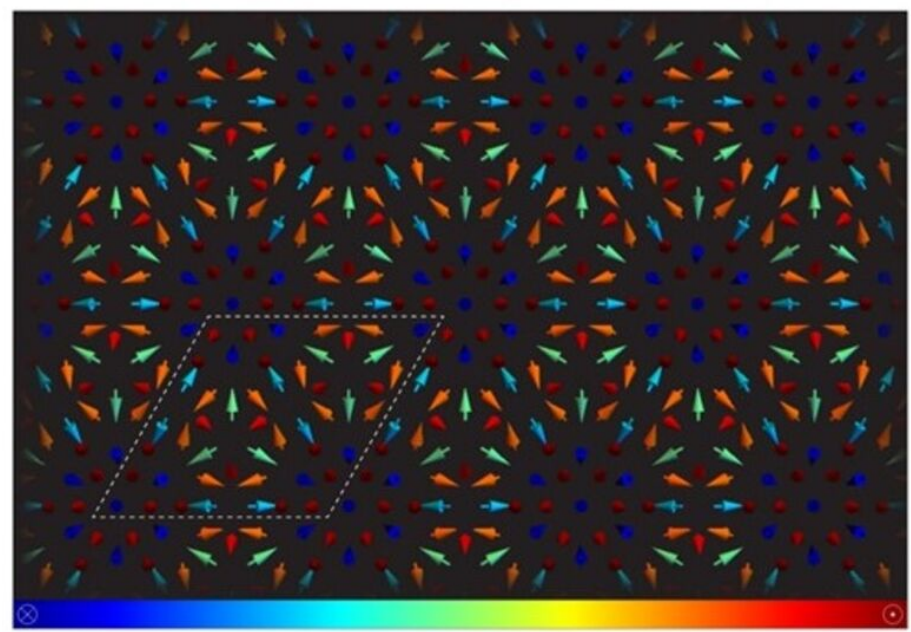 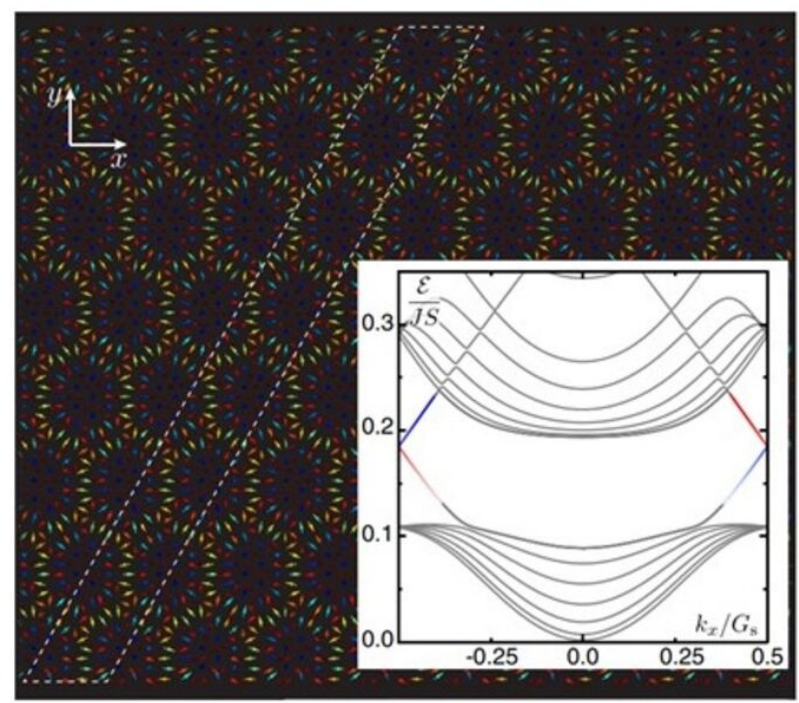 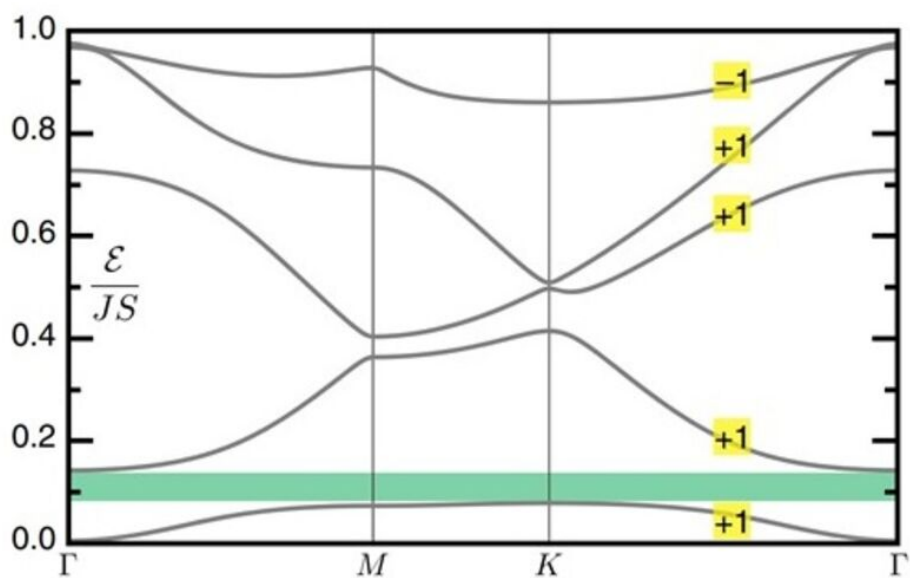 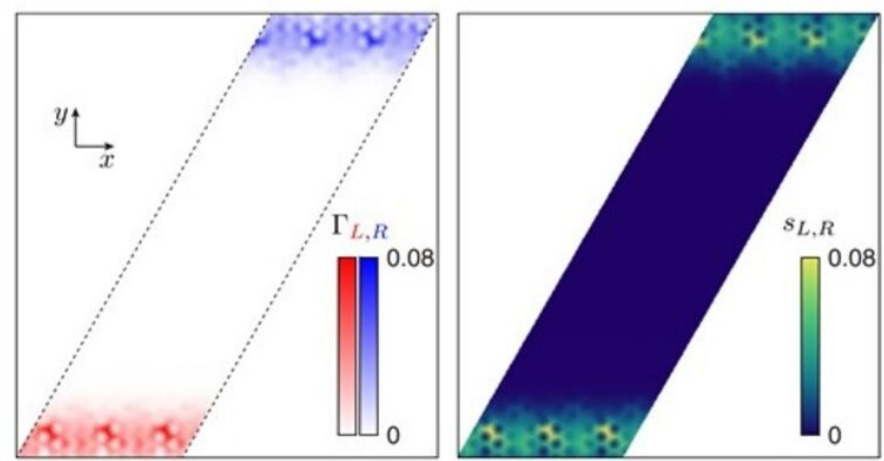

**(b)** 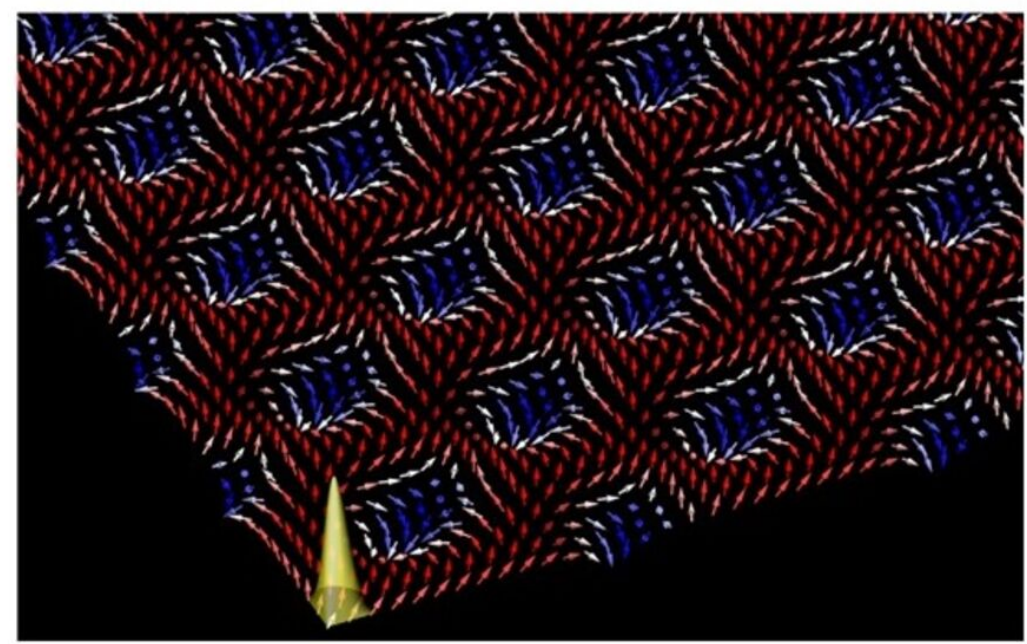 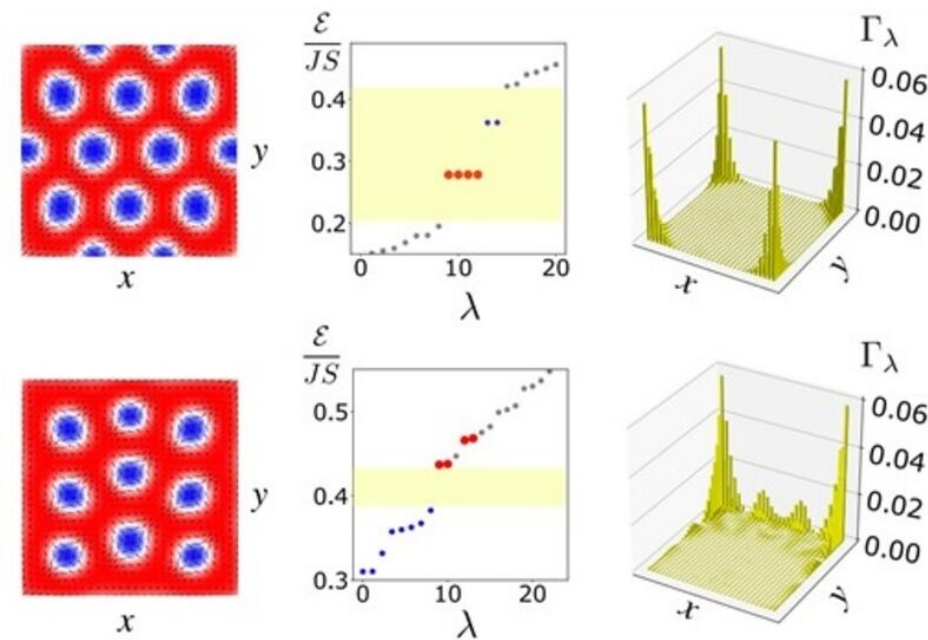 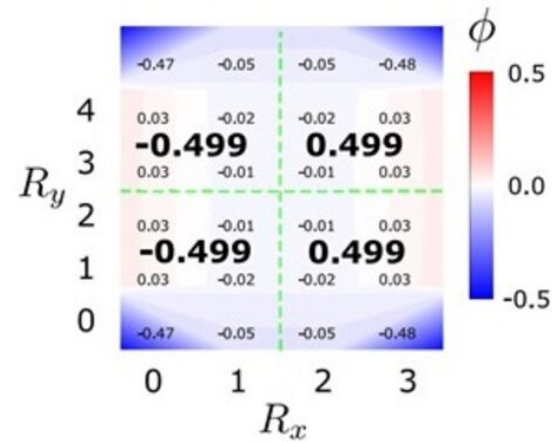 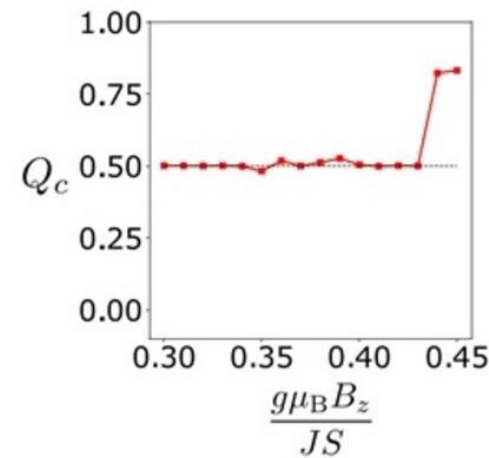 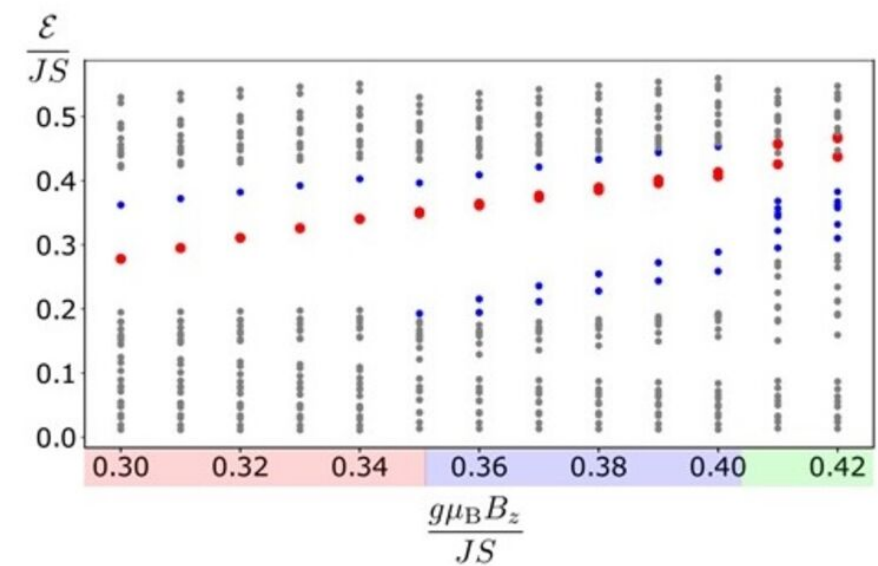

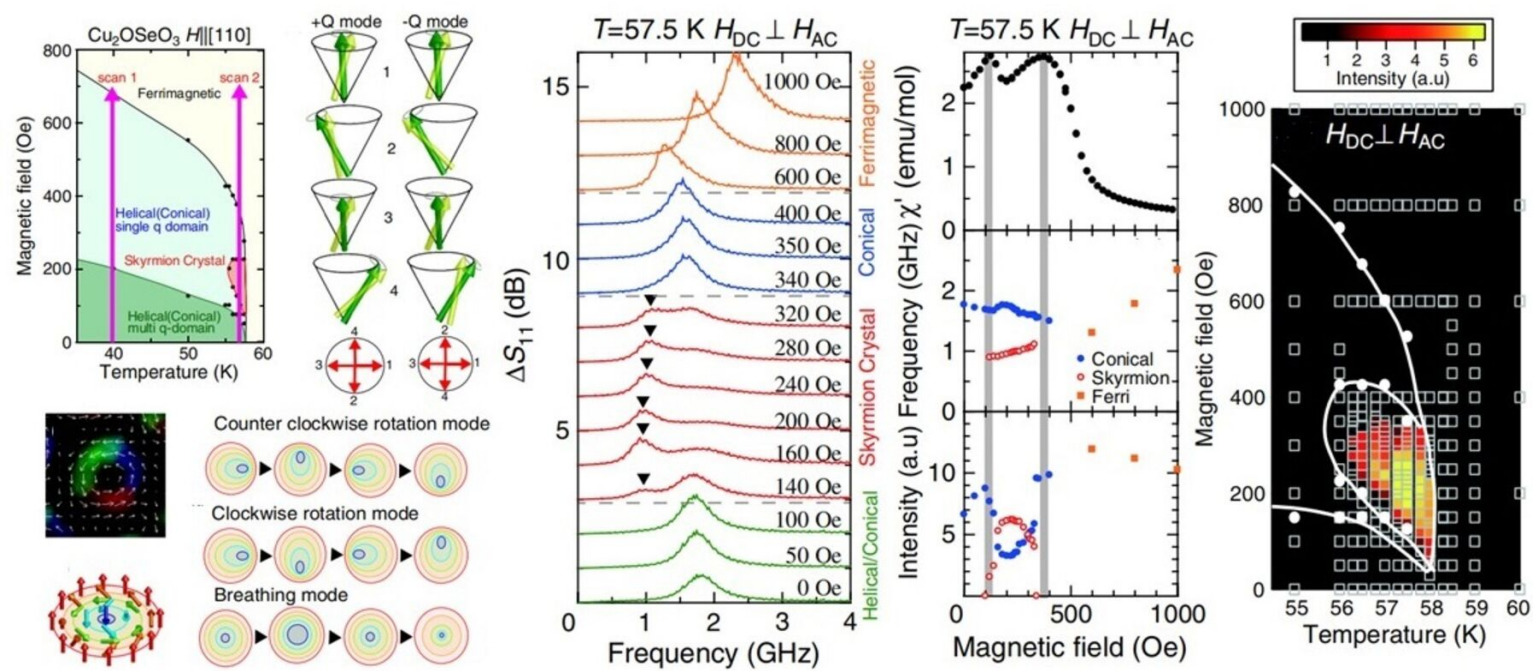
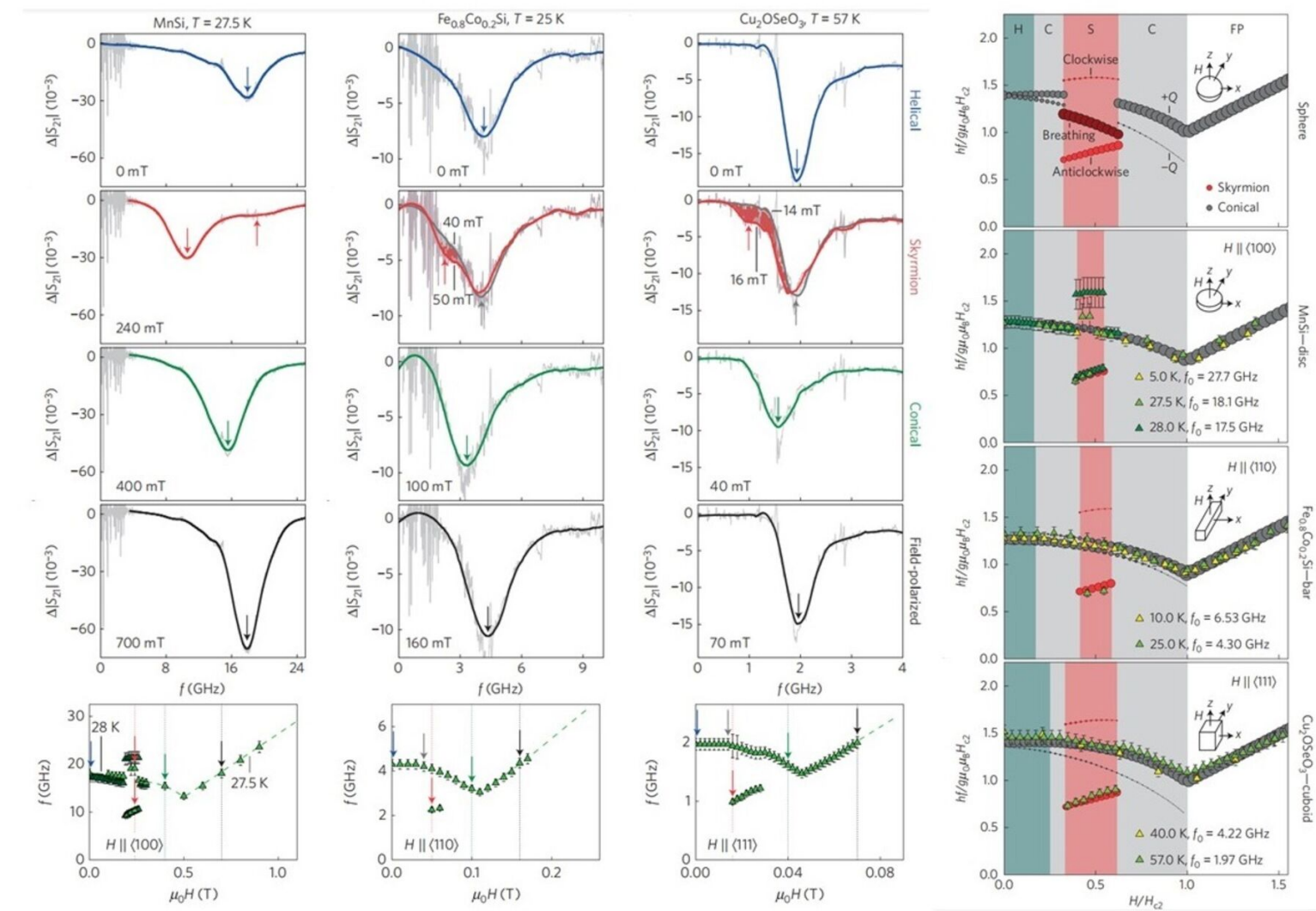
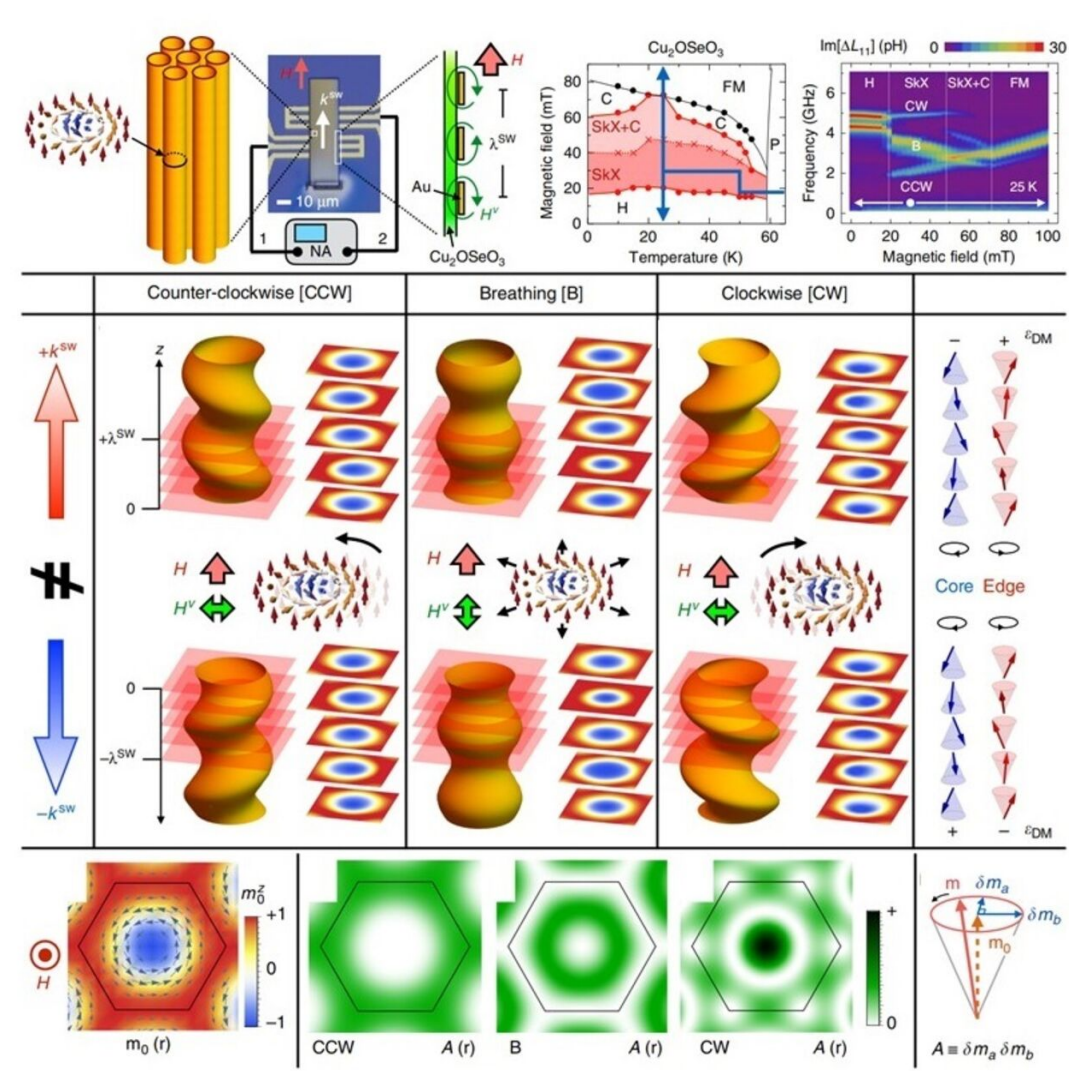
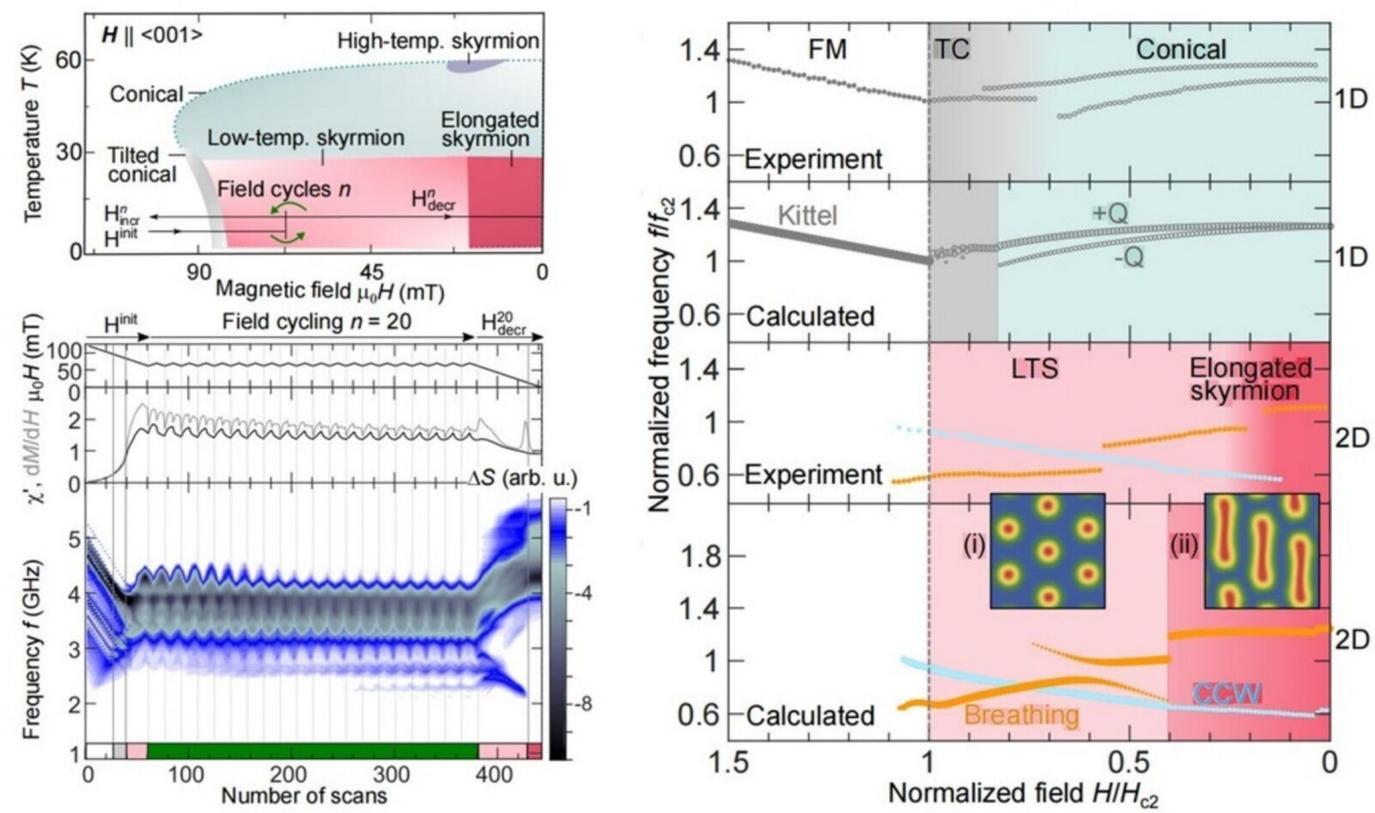

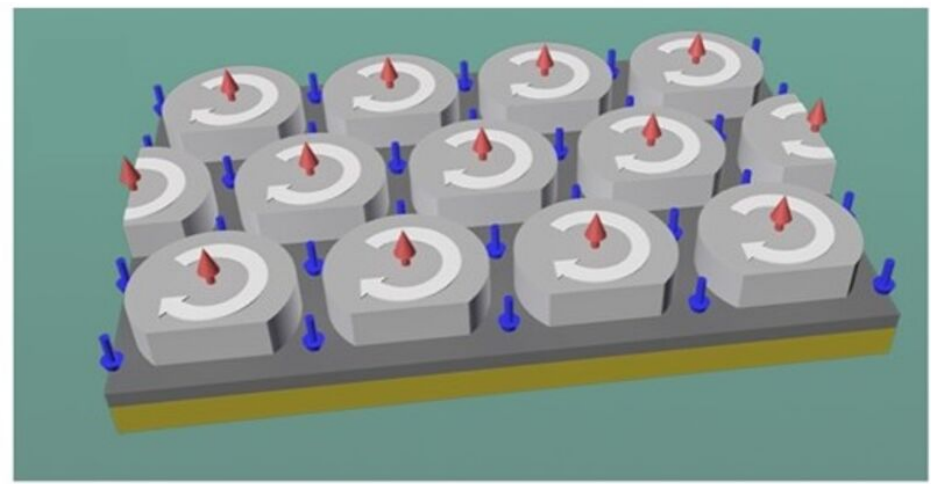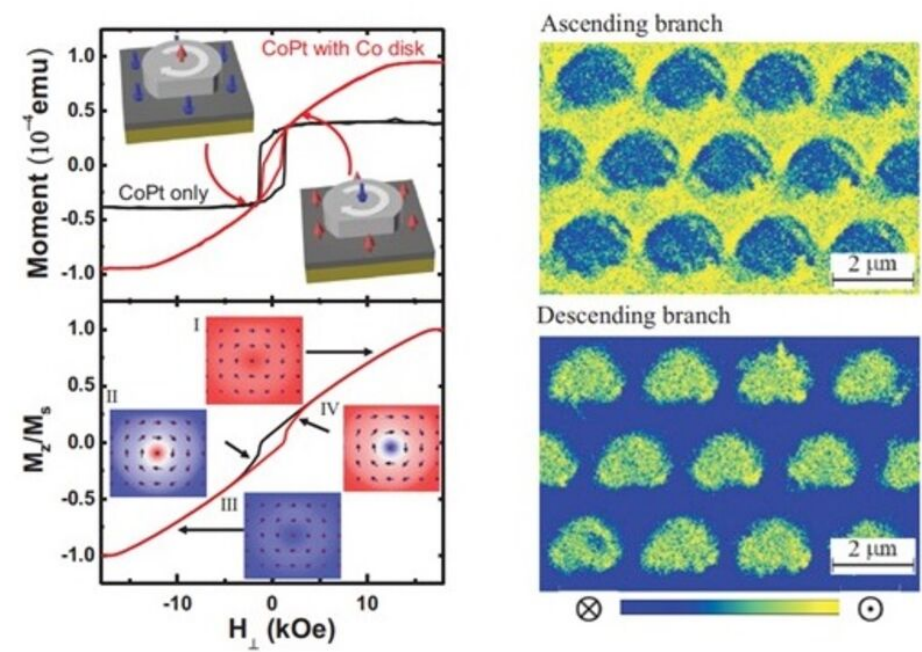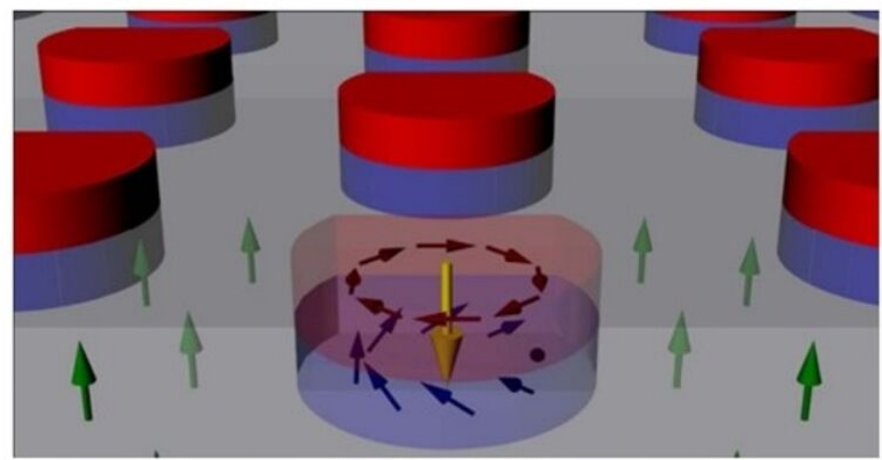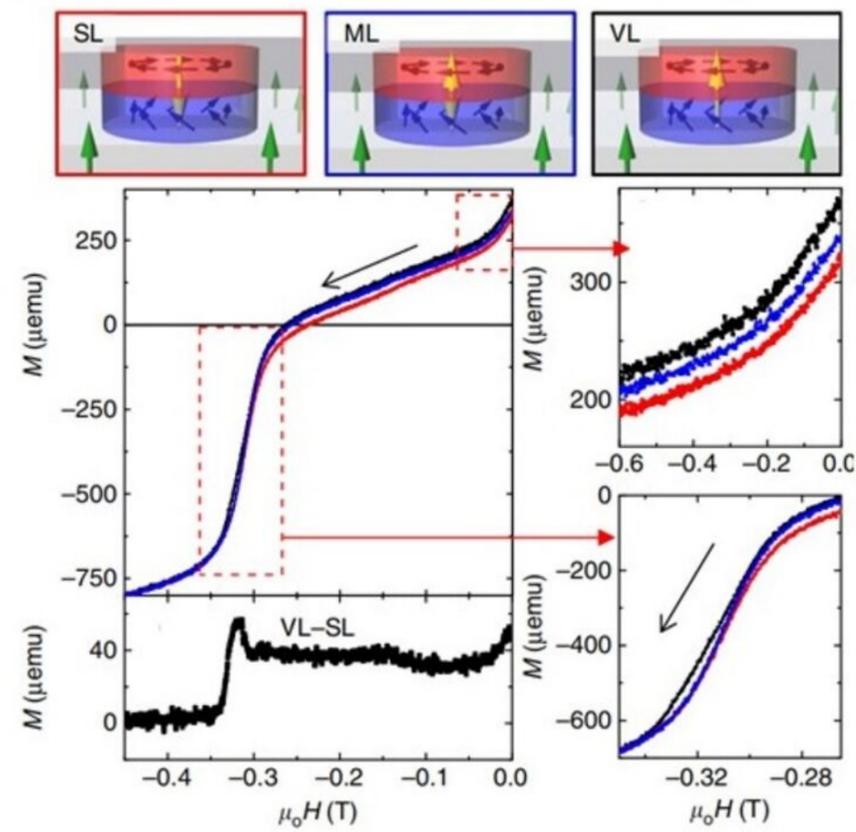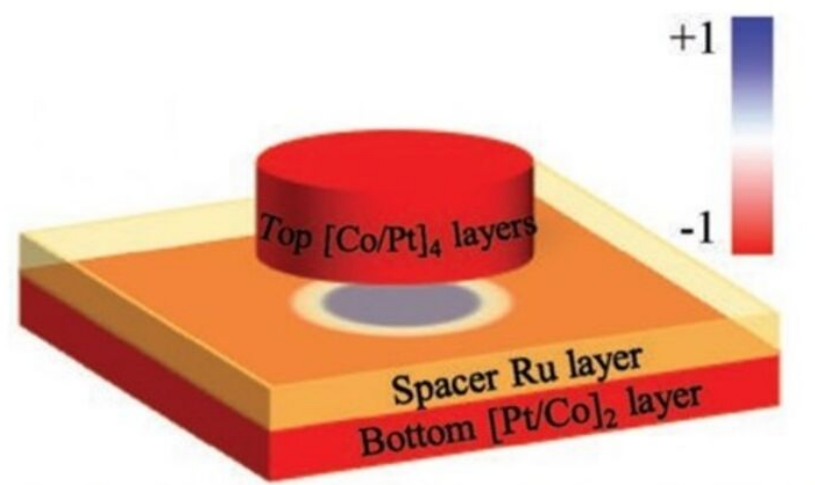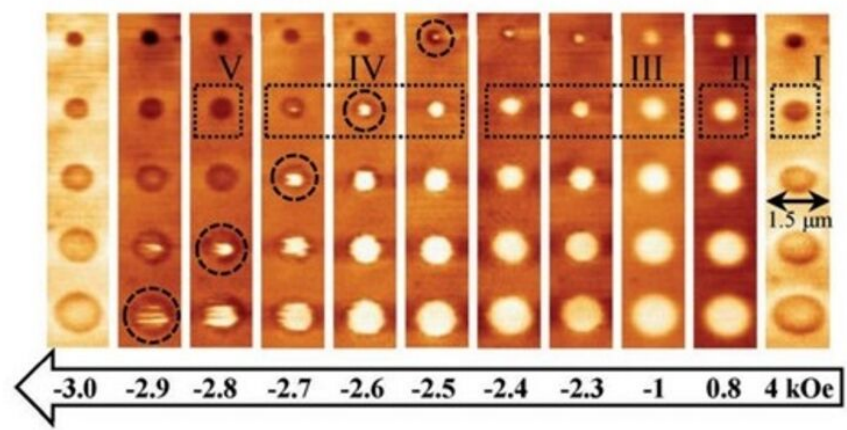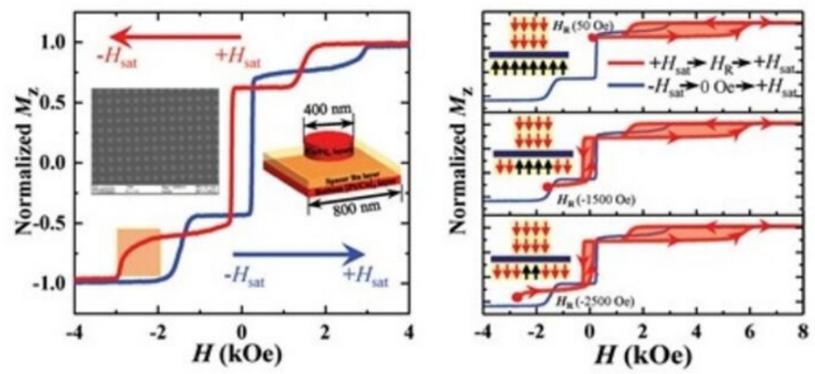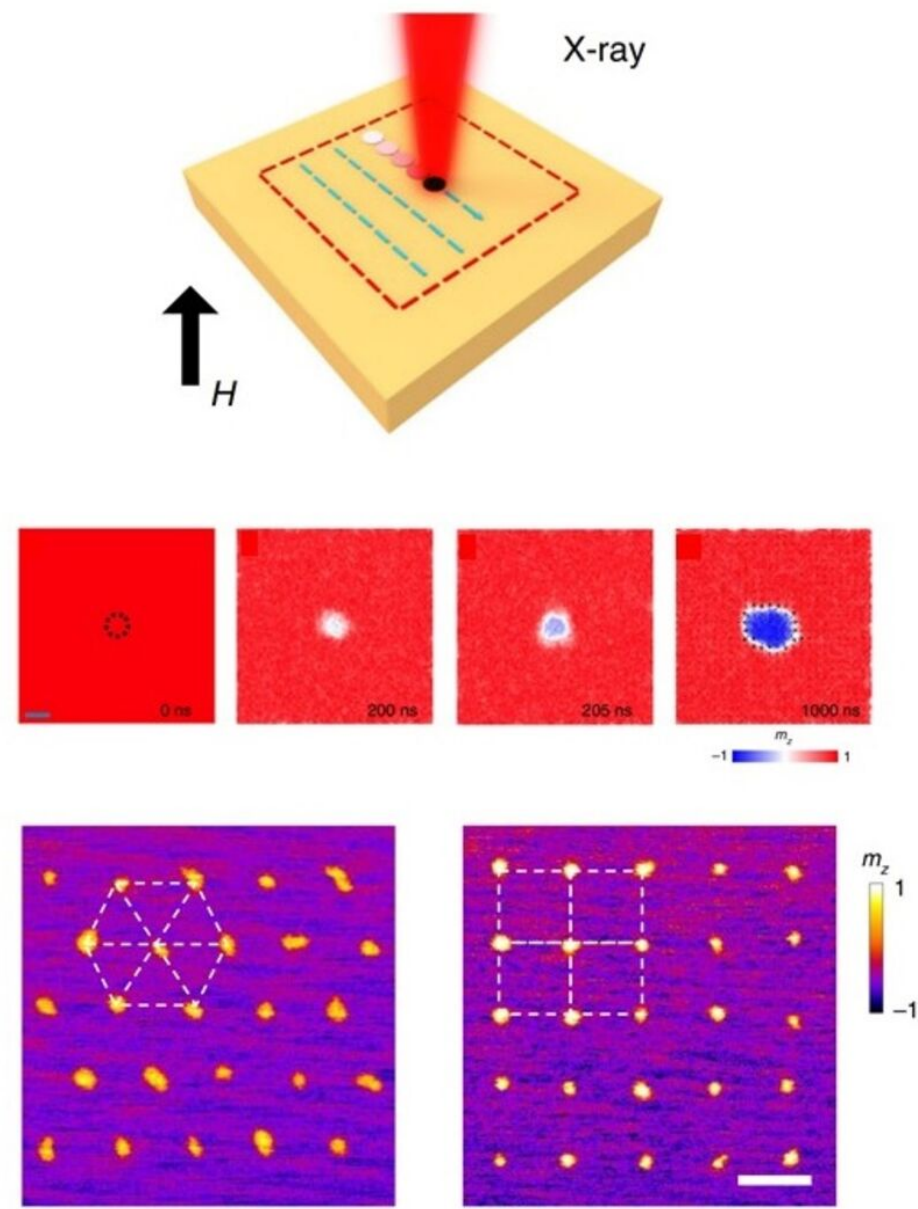

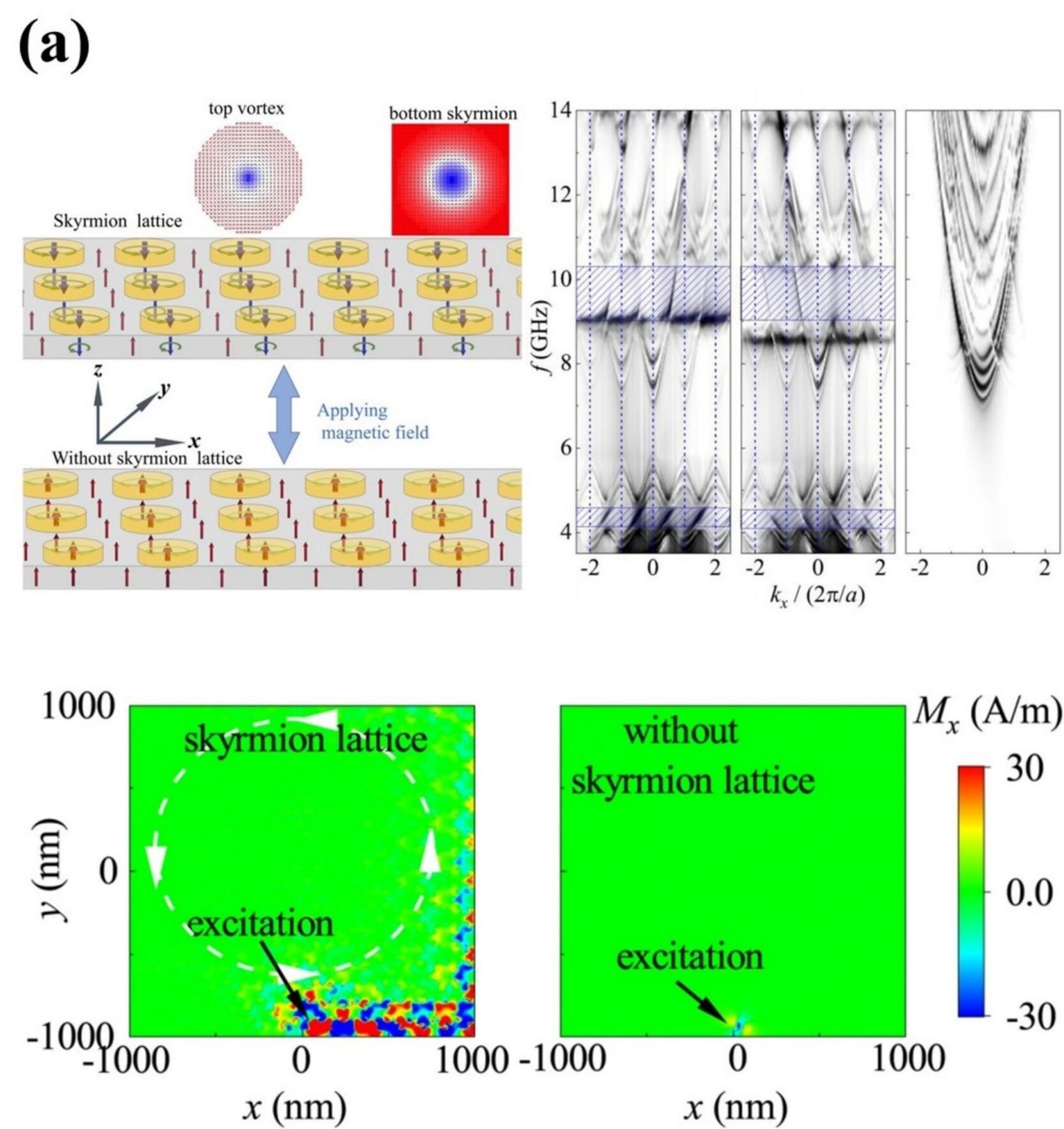
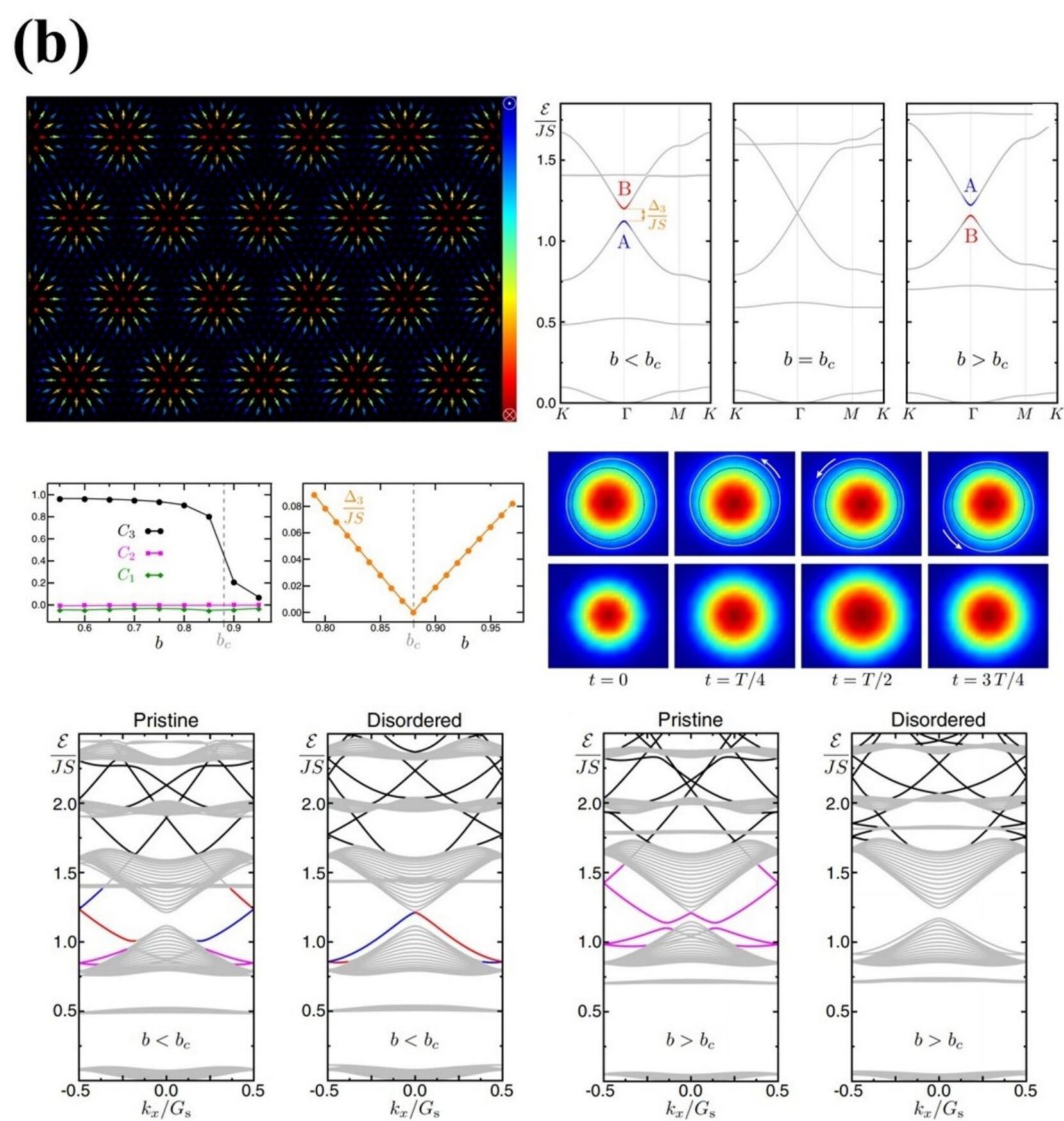

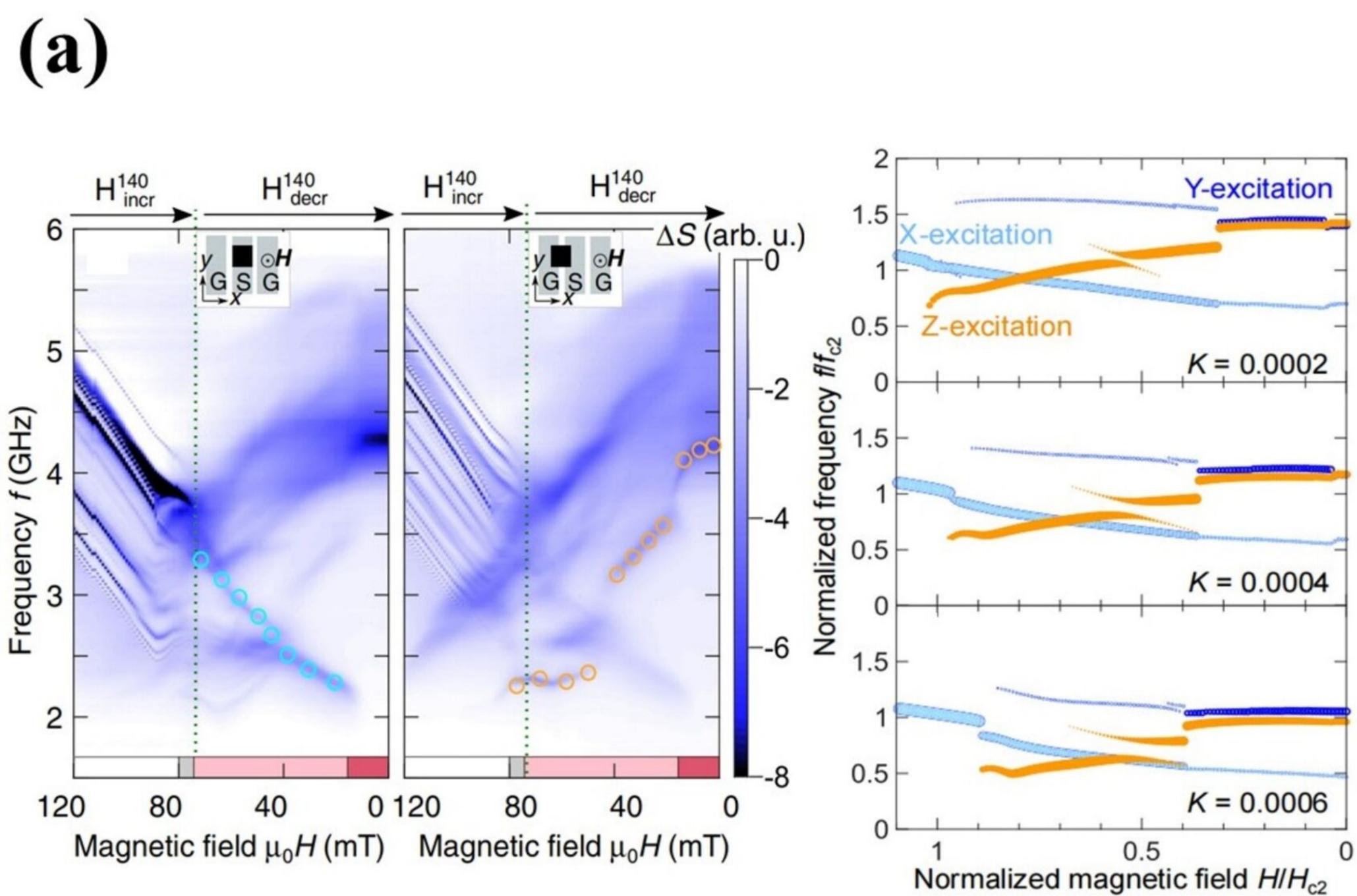
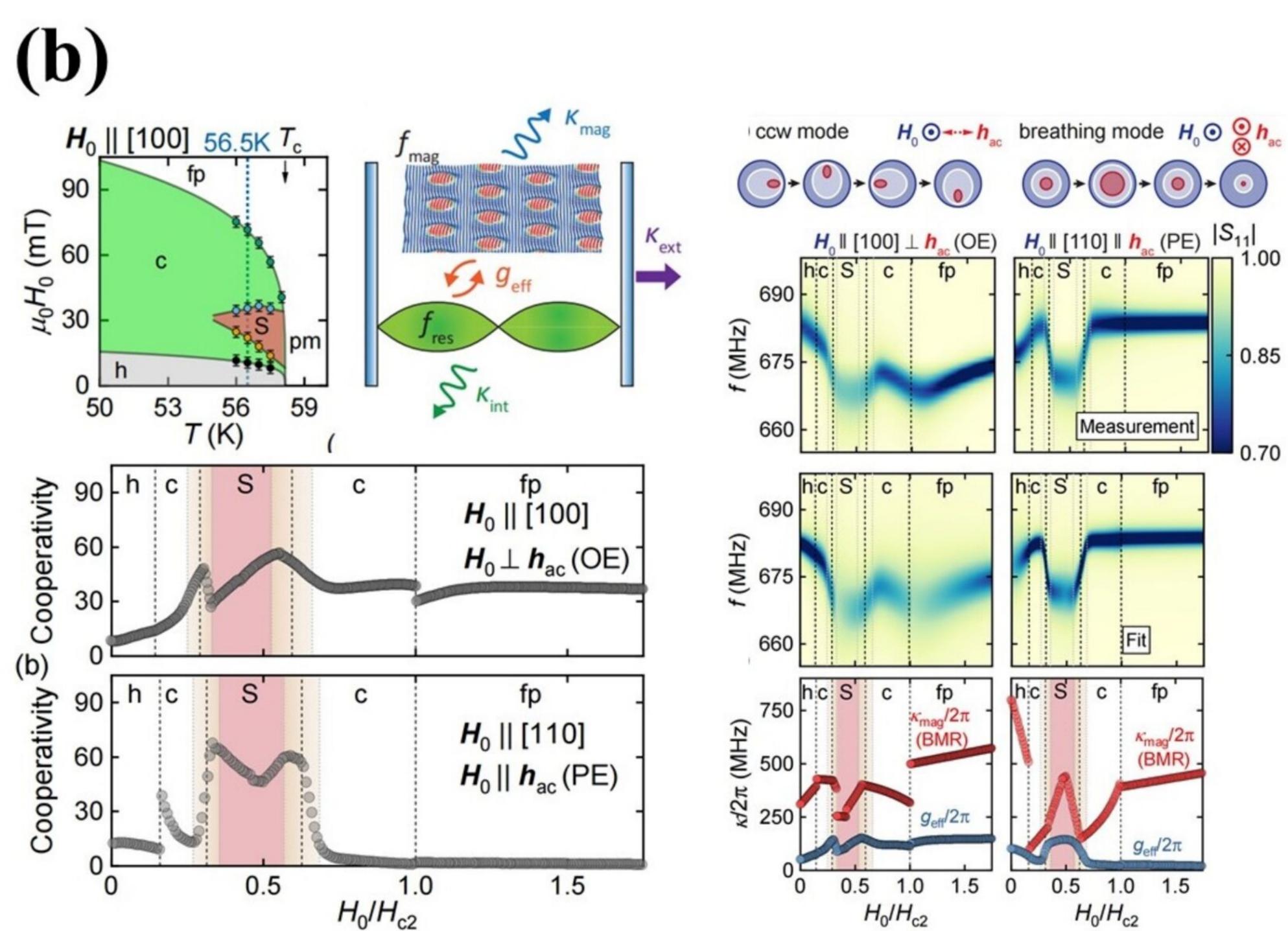